\documentclass[fleqn,usenatbib]{mnras}

\usepackage[T1]{fontenc}
\usepackage{ae,aecompl}

\usepackage{graphicx}	% Including figure files
\usepackage{amsmath}	% Advanced maths commands
\usepackage{amssymb}	% Extra maths symbols
\usepackage{xcolor}
\usepackage{booktabs}
\hypersetup{draft}
\definecolor{gray}{rgb}{0.0, 0.42, 0.24}
\definecolor{orange}{rgb}{0.82, 0.1, 0.26}
\usepackage{threeparttable}

\title[Dust-gas chemistry in AGB outflows]{Chemical modelling of dust-gas chemistry within AGB outflows \\ II. Effect of the dust-grain size distribution}

\author[Van de Sande et al.]{
M. Van de Sande$^{1}$\thanks{E-mail: marie.vandesande@kuleuven.be}\thanks{Postdoctoral Fellow of the Fund for Scientific Research (FWO), Flanders, Belgium},
C. Walsh$^{2}$,
T. Danilovich$^{1}$
\\
$^{1}$Department of Physics and Astronomy, Institute of Astronomy, KU Leuven, Celestijnenlaan 200D, 3001 Leuven, Belgium\\
$^{2}$School of Physics and Astronomy, University of Leeds, Leeds LS2 9JT, UK\\
}

\date{Accepted XXX. Received YYY; in original form ZZZ}

\pubyear{2015}

\begin{document}
\label{firstpage}
\pagerange{\pageref{firstpage}--\pageref{lastpage}}
\maketitle

\begin{abstract}
AGB stars are, together with supernovae, the main contributors of stellar dust to the interstellar medium (ISM).
Dust grains formed by AGB stars are thought to be large.
However, as dust nucleation and growth within their outflows are still not understood, the dust-grain size distribution (GSD) is unknown.
This is an important uncertainty regarding our knowledge of the chemical and physical history of interstellar dust, as AGB dust forms $\sim 70\%$ of the starting point of its evolution.
We expand on our chemical kinetics model, which uniquely includes a comprehensive dust-gas chemistry.
The GSD is now allowed to deviate from the commonly assumed canonical \citet{Mathis1977} distribution.
We find that the specific GSD can significantly influence the dust-gas chemistry within the outflow.
Our results show that the level of depletion of gas-phase species depends on the average grain surface area of the GSD.
Gas-phase abundance profiles and their possible depletions can be retrieved from observations of molecular emission lines when using a range of transitions.
Due to degeneracies within the prescription of GSD, specific parameters cannot be retrieved, only (a lower limit to) the average grain surface area.
Nonetheless, this can discriminate between dust composed of predominantly large or small grains.
We show that when combined with other observables such as the spectral energy distribution and polarised light, depletion levels from molecular gas-phase abundance profiles can constrain the elusive GSD of the dust delivered to the ISM by AGB outflows.
\end{abstract}


%%%%%%%%%%%%%%%%%%%%%%%%%%%%%%%%%%%%%%%%%%%%%%%%%%

%%%%%%%%%%%%%%%%% BODY OF PAPER %%%%%%%%%%%%%%%%%%

%%%%%%%%%%%%%%%%%%%%%%%%%%%%%%%%%%%%%%%%%%%%%%%%%%%%%%%%%%%%%%%%%%%%%%%%%%%%%%%%%%%%%%%%%%%%%%%%%%%%%%%%%%%%%%
\section{Introduction}
%%%%%%%%%%%%%%%%%%%%%%%%%%%%%%%%%%%%%%%%%%%%%%%%%%%%%%%%%%%%%%%%%%%%%%%%%%%%%%%%%%%%%%%%%%%%%%%%%%%%%%%%%%%%%%

Dust plays an important role in the interstellar medium (ISM). 
Dust grains absorb optical and ultraviolet (UV) radiation and re-emit it in the infrared, affecting the spectral energy distribution and energy balance of the environment. 
They facilitate the formation of molecules, especially H$_2$, by providing the catalytic surfaces for surface chemistry \cite[e.g.,][]{Gould1963,Cazaux2004}.
Both the radiative properties and the H$_2$ formation rate depend on the specific dust-grain size distribution (GSD) \cite[e.g.,][]{Takeuchi2003}.
Different prescriptions exist for the GSD of interstellar dust, derived from extinction curves and other observations, such as polarisation and spectroscopy \cite[see e.g.,][]{Mathis1977,Li2001b,Cecchi-Pestellini2010,Jones2013,Williams2016}.
The GSD is the result of the evolution of dust in the ISM, where it undergoes accretion, coagulation, shattering, sputtering and thermal processing, which are all time- and space-dependent processes \cite[e.g.,][]{Draine1979,Dwek1980,Jones1996,Dwek1998,Yan2004,Ormel2009,Asano2013,Hirashita2015}.
To fully understand the chemical and physical history of interstellar dust -- encrypted in its GSD -- the input of stellar dust needs to be accurately quantified, as it forms the starting point of the dust evolution cycle in the ISM.

Asymptotic giant branch (AGB) stars and supernovae are the main contributors of stellar dust to the ISM, contributing $\sim 70\%$ of the total stardust production rate \citep{Zhukovska2013}.
During the AGB phase, stars of low-to-intermediate mass lose their outer layers by means of a stellar wind or outflow, creating an extended circumstellar envelope (CSE).
The wind is thought to be driven by a two-step mechanism: stellar pulsations facilitate the formation of dust grains, which subsequently leads to a dust-driven outflow \citep{Hofner2018}. 
Observations of {spectral energy distributions (SEDs)}, theoretical studies, as well as meteoric samples suggest that the typical grain size of AGB dust is large, $a \geq 0.1\ \mu$m ($a \geq 10^{-5}$ cm), but not single-sized \citep{Groenewegen1997,Winters1997,Gauger1999,Hoppe2000,Yasuda2012,DellAgli2017,Nanni2018}.
Measurements of the grain size close to the star through polarisation show that large grains ($a \sim 0.3\ \mu$m) can be formed close to O-rich AGB stars \citep{Norris2012}, {although the presence of smaller grains ($a < 0.1\ \mu$m) in this region is also suggested \citep{Khouri2020}}.
However, such measurements are difficult due to the strong contamination by molecular bands that disrupt the wavelength dependence imprinted by scattering \citep{Khouri2016}.
Radiation-hydrodynamical models have also suggested that radiation pressure on micron-sized Fe-free silicates close to the star is a potential launching mechanism of O-rich winds \citep{Hofner2008}.

\begin{table}
	\caption{Physical parameters of the grid of chemical models.}
	\resizebox{1.0\columnwidth}{!}{%
%	\centering
	\label{table:modelparams}
	\begin{tabular}{ll} % four columns, alignment for each
		\hline
    Mass-loss rate, $\dot{M}$        &    $10^{-5}, 10^{-6} \ \mathrm{M}_\odot\ \mathrm{yr}^{-1}$ \\
    Outflow velocity, $v_\infty$   & 5, 15 km s$^{-1}$  \\
    Stellar temperature, $T_*$        & 2000 K \\
    Stellar radius, $R_*$             & 5 $\times 10^{13}$ cm \\
    Exponent temperature power-law, $\epsilon$                       & 0.7 \\
    Drift velocity, $v_\mathrm{drift}$	& 5, 10, 15 km s$^{-1}$  \\
    Initial radius of the model		& $10^{15}$ cm \\
    Final radius of the model		& $10^{18}$ cm \\
		\hline
	\end{tabular}
	}
\end{table}

Exactly how the dust is formed and grows is still unknown both theoretically and observationally. 
Therefore, the GSD of dust formed in AGB outflows is largely unknown, despite the extensive modelling of the dust shells around specific stars, as well as the modelling that included the contribution of AGB dust to the dust production rate of galaxies \cite[e.g.,][]{Groenewegen1997,Winters1997,Gauger1999,Hofner2008,Zhukovska2013,Khouri2016,DellAgli2017,Nanni2018}.
Besides its importance to the evolution of dust in the ISM, this also influences the study of AGB outflows themselves.
Single-sized grains and the canonical \citet{Mathis1977} (MRN) distribution are commonly assumed when modelling CSEs, despite the influence of the GSD on the dust temperature and overall energy balance of the CSE, and hence the relative emission strength of molecular lines.
Accurate retrieval of the physical and chemical properties of AGB outflows therefore depends on the use of a realistic GSD.

\citet{VandeSande2019b} (henceforth Paper I) extended a gas-phase only chemical kinetics model of a CSE to include dust-gas interactions and surface chemistry.
They found that dust-gas chemistry can cause significant depletions of gas-phase species, covering the dust grains with an ice mantle.
The level of depletion depends on the dust grain temperature profile and the outflow density.
In Paper I, the canonical MRN distribution was used as the GSD.
In this paper, we assess the dependency of the level of gas-phase depletion on the assumed GSD and whether molecular line emission can be used as a tracer of the GSD present within the outflow.

The chemical kinetics and radiative transfer models are described in Sect. \ref{sect:meth}, together with a description of the parameter selection.
The results of the influence of the GSD on the gas-phase abundances and the molecular line emission are shown in Sect. \ref{sect:results}.
They are discussed in Sect. \ref{sect:discussion}, followed by the conclusions in Sect. \ref{sect:conclusions}.

%%%%%%%%%%%%%%%%%%%%%%%%%%%%%%%%%%%%%%%%%%%%%%%%%%%%%%%%%%%%%%%%%%%%%%%%%%%%%%%%%%%%%%%%%%%%%%%%%%%%%%%%%%%%%%
\section{Methodology}				\label{sect:meth}
%%%%%%%%%%%%%%%%%%%%%%%%%%%%%%%%%%%%%%%%%%%%%%%%%%%%%%%%%%%%%%%%%%%%%%%%%%%%%%%%%%%%%%%%%%%%%%%%%%%%%%%%%%%%%%

The chemical model is described in Sect. \ref{subsect:meth:chemistry}, where we elaborate on corrections to the model of Paper I, and describe the inclusion of the GSD in the model and the parameter selection in detail.
The radiative transfer model used to extract observables from the calculated abundance profiles is described in Sect. \ref{subsect:meth:rt}.

%-------------------------------------------------------------------------------------------------------------
\subsection{Chemical model}			\label{subsect:meth:chemistry}
%-------------------------------------------------------------------------------------------------------------

\begin{table}
	\caption{Parent species for the C-rich and O-rich CSE and their initial abundances relative to H$_2$ and binding energies $E_\mathrm{bind}$. 
} 
    \centering
    \begin{tabular}{l r l c}
    \hline  
    % \noalign{\smallskip}
    \multicolumn{4}{c}{Carbon-rich}   \\  
    \cline{1-4}
    \noalign{\smallskip}
    Species & Abundance & $E_\mathrm{bind}$ (K) & Ref.  \\
    \hline
    % \noalign{\smallskip}
    He            & 0.17                 & 100 &        \\
    CO            & $8.0\times10^{-4}$   & 855 & (1)     \\
    N$_2$         & $4.0\times10^{-5}$   & 790 & (2)     \\
    C$_2$H$_2$    & $8.0\times10^{-5}$   & 2090 & (3)    \\
    HCN           & $2.0\times10^{-5}$   & 3610 & (3)    \\
    SiO           & $1.2\times10^{-7}$   & 3500 & (3)    \\
    SiS           & $1.0\times10^{-6}$   & 3800 & (3)     \\
    CS            & $5.0\times10^{-7}$   & 1900 & (3)      \\
    SiC$_2$       & $5.0\times10^{-8}$   & 1300 & (3)     \\
    HCP           & $2.5\times10^{-8}$   & 1100 & (3)    \\
    NH$_3$        & $2.0\times10^{-6}$   & 2715 & (4)     \\
    H$_2$O        & $1.0\times10^{-7}$   & 4880 & (5)    \\
    \hline 
    \multicolumn{4}{c}{Oxygen-rich}   \\  
    \cline{1-4}
    \noalign{\smallskip}
    Species & Abundance & $E_\mathrm{bind}$ (K) & Ref.  \\
    \hline
    He        & 0.17                & 100 &     \\
    CO        & $3.0\times10^{-4}$  & 855 & (1) \\
    N$_2$     & $4.0\times10^{-5}$  & 790 & (2) \\
    H$_2$O    & $3.0\times10^{-4}$  & 4880 & (6)  \\
    CO$_2$    & $3.0\times10^{-7}$  & 2267 & (7)  \\
    SiO       & $5.0\times10^{-5}$  & 3500 & (8)  \\
    SiS       & $2.7\times10^{-7}$  & 3800 & (9)  \\
    SO        & $1.0\times10^{-6}$  & 1800 & (10)  \\
    H$_2$S    & $7.0\times10^{-8}$  & 2290 & (11)  \\
    PO        & $9.0\times10^{-8}$  & 1150 & (12)  \\
    HCN       & $2.0\times10^{-7}$  & 3610 & (13)  \\
    NH$_3$    & $1.0\times10^{-7}$  & 2715 & (14) \\
    \hline 
    \end{tabular}%
    % }
    \\
    \footnotesize
    { {{References.}} (1) \citet{Teyssier2006}; (2) TE abundance \citep{Agundez2010}; (3) \citet{Agundez2009};
    (4) \citet{Agundez2012}; (5) \citet{Decin2010b}; (6) \citet{Maercker2008}; (7) \citet{Tsuji1997};
    (8) \citet{Schoier2004}; (9) \citet{Schoier2007}; (10) \citet{Bujarrabal1994};
    (11) \citet{Ziurys2007}; (12) \citet{Tenenbaum2007}; (13) \citet{Decin2010}; (14) \citet{Wong2018}.
    }
    \label{table:Model-Parents}    
\end{table}

\begin{table}
	\centering
	\caption{Parameters of the dust-grain size distribution. The upper three parameters of the MRN-like GSD are varied throughout the modelling, while maintaining $a_\mathrm{min} < a_\mathrm{max}$. The lower parameters are fixed.}
	\label{table:dustparams}
	\begin{tabular}{lr} % four columns, alignment for each
		\hline
	Minimum grain size, $a_\mathrm{min}$ &  \\
	\multicolumn{2}{r}{$10^{-8}$, $5 \times 10^{-7}$, $10^{-7}$, $10^{-6}$ cm} \\
	Maximum grain size, $a_\mathrm{max}$ & \\
	\multicolumn{2}{r}{$10^{-7}$, $10^{-6}$, $2.5 \times 10^{-5}$, $10^{-5}$, $10^{-4}$ cm} \\
	Exponent, $\eta$ & \\
	\multicolumn{2}{r}{$-5.5, -4.5, -3.5, -2.5, -1.5, -0.5, +0.5, +1.5$} \\
	\hline
	Dust-to-gas mass ratio, $\psi$			& $2 \times 10^{-3}$ \\
	Surface density of binding sites, $n_s$	& $10^{15}$ cm$^{-2}$ \\
	Silicate dust bulk density$^1$, $\rho_\mathrm{dust,bulk}$ 	& 3.5 g cm$^{-3}$  \\
	Carbonaceous dust bulk density$^1$, $\rho_\mathrm{dust,bulk}$ 	& 2.24 g cm$^{-3}$  \\
		\hline
	\end{tabular}
    \footnotesize
    { {{References.}} (1) \citet{Draine2003}
    }
\end{table}

The chemical kinetics model and reaction network used are those of Paper I.
This one-dimensional model, based on the publicly available UMIST Database for Astrochemistry (UDfA) CSE model \citep{McElroy2013}\footnote{\url{http://udfa.ajmarkwick.net/index.php?mode=downloads}}, describes a uniformly expanding outflow with a constant mass-loss rate and outflow velocity.
{The gas temperature throughout the outflow is assumed to follow a power law,
\begin{equation}
	T_\mathrm{gas}(r) = T_* \left( \frac{r}{R_*} \right)^{-\epsilon},
\end{equation}
where $T_*$ and $R_*$ are the stellar temperature and radius, respectively, and $r$ is the distance from the centre of the star.}

The reaction network is an extension of the gas-phase only \textsc{Rate12} network \citep{McElroy2013} and includes a comprehensive dust-gas chemistry: dust and gas can interact through accretion (forming an ice mantle) and thermal desorption, photodesorption, and sputtering (destroying the ice mantle).
Chemical reactions can occur on the surface of the dust via both the diffusive Langmuir-Hinshelwood and the stick-and-hit Eley-Rideal mechanisms. 
A complete description of the dust-gas chemistry is given in Paper I.
We assume that dust grains are present throughout the outflow.
Our models start at $10^{15}$ cm $\sim$ 20 R$_*$ from the stellar surface; hence, the dust is assumed to have formed with a specific GSD in the region within 20 R$_*$.
We assume that the GSD does not change throughout the outflow. 
Gas-phase species are able to form an ice mantle surrounding the grain through accretion, but they are not chemically incorporated into the dust.

The physical parameters for the models considered in this paper are given in Table \ref{table:modelparams}.
They correspond to the outflows for which significant levels of depletion of gas-phase parent species onto the dust were found in Paper I, namely higher density outflows, with $\dot{M} = 10^{-5}$ M$_\odot$ yr$^{-1}$ and $v_\infty = 5$ and 15 km s$^{-1}$ and $\dot{M} = 10^{-6}$ M$_\odot$ yr$^{-1}$ and $v_\infty = 5$ km s$^{-1}$. 
{Although lower density outflows are common \cite[e.g.,][]{Danilovich2015}, we do not consider low-density outflows here, as they do not show depletion of gas-phase species onto the dust (see Paper I).}
The highest density outflow  investigated is not common around AGB stars \cite[e.g.,][]{Danilovich2015}.
However, we have included it to sample the entire physical parameter space.
The drift velocity $v_\mathrm{drift}$ is varied over 5, 10 and 15 km s$^{-1}$. 
All other parameters are kept constant.
Table \ref{table:Model-Parents} lists the parent species, together with their relative abundances and binding energies, for the O-rich and C-rich outflows.
These values were also used in Paper I.

{The temperature of the dust grains throughout the outflow depends on the dust type and the outflow density.
The dust temperature profile is approximated by a power law,
\begin{equation}        \label{eq:tdust}
    T_\mathrm{dust}(r) = T_\mathrm{dust,*} \left( \frac{2r}{R_*} \right)^{-\frac{2}{4+s}},
\end{equation}
where $T_\mathrm{dust,*}$ and $s$ are free parameters. 
These are obtained from fitting Eq. (\ref{eq:tdust}) to the results of the continuum radiative transfer code \textsc{MCMax} \citep{Min2009}.
We use the results of Paper I; the values of $T_\mathrm{dust,*}$ and $s$ are listed in table 4 of that paper.
The dust is assumed to be composed of a single species to reduce the number of free parameters.
For the O-rich outflows, we consider three types of dust. 
Each has a different temperature profile, which has a large influence on the depletion of gas-phase species (see Paper I).
The species are, in order of decreasing dust temperature}, MgFeSiO$_4$ (olivine with iron; optical constants from \citeauthor{Jager1994}~\citeyear{Jager1994}),  Mg$_2$SiO$_4$ (olivine without iron; \citeauthor{Jager2003}~\citeyear{Jager2003}) and Ca$_2$Mg$_{0.5}$Al$_2$Si$_{1.5}$O$_7$ (melilite; \citeauthor{Mutschke1998}~\citeyear{Mutschke1998}).
{Modelling of SEDs shows that these types of silicates are common around O-rich AGB stars \cite[e.g.,][]{Heras2005}.}
For the C-rich outflows, we consider amorphous carbon only (optical constants from \citeauthor{Preibisch1993}~\citeyear{Preibisch1993}).
The optical constants {used in the continuum radiative transfer code} were calculated for particle shapes represented by a distribution of hollow spheres with a filling factor of 0.8 \citep{Min2003}. Olivine with and without iron had $a_\mathrm{min}$ = 0.01 $\mu$m and $a_\mathrm{max}$ = 3 $\mu$m, melilite had $a_\mathrm{min}$ = 0.29 $\mu$m and $a_\mathrm{max}$ = 0.31 $\mu$m.
For amorphous carbon, the opacities were calculated using both the continuous distribution of ellipsoids approximation, for grains of 1 $\mu$m in size.

{
We acknowledge that the GSDs encoded in the optical data, used only in the continuum radiative transfer code for computation of the dust temperature, differs from those investigated in the chemical models.}
This is because these are the only available optical data in the Leuven version of \textsc{MCMax} for the different dust types investigated.
Running radiative transfer models with GSDs that are consistent with those used in the chemical model would require the bespoke computation of optical constants and is beyond the scope of this work.
Note that the variation in the chemical composition of the dust appears to have a larger influence on the resulting dust temperature profiles than the assumed GSD (see Paper I).
{The difference between the GSDs used in the continuum radiative transfer code and those used in the chemical model (see Sect. \ref{subsubsect:gsd}) are therefore not expected to have a large impact.}

\begin{figure*}
 \includegraphics[width=0.90\textwidth]{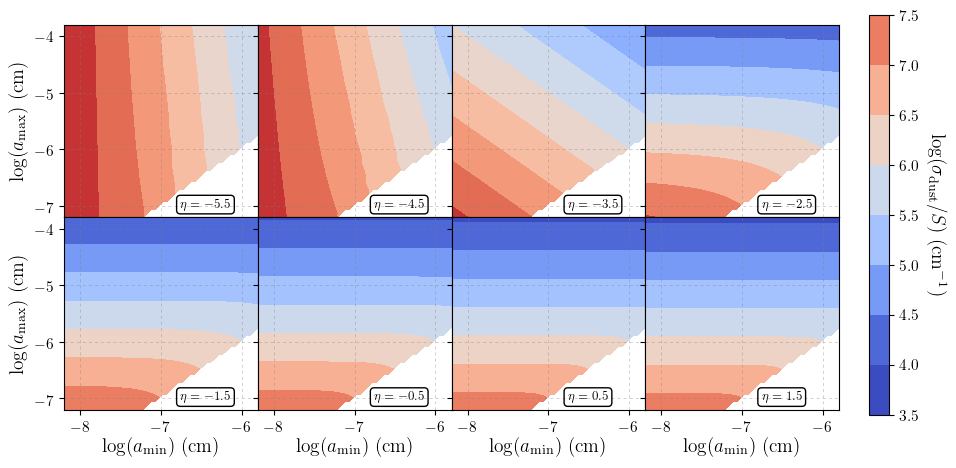}
 \caption{Dependence of $\sigma_\mathrm{dust}/S$ on the parameters $a_\mathrm{min}$, $a_\mathrm{max}$, and $\eta$, varied over the values listed in Table \ref{table:dustparams}. 
 The parameter $\sigma_\mathrm{dust}/S$ (Eq. \ref{eq:sigmadust}) is an outflow-independent proxy for the average dust grain cross section per unit volume and includes the three parameters of the MRN-like GSD, $a_\mathrm{min}$, $a_\mathrm{max}$, and $\eta$, the slope of the GSD (Eq. \ref{eq:ndust}).
 The different panels correspond to different values of $\eta$.
  }
 \label{fig:sigma}
\end{figure*}

%--.--.--.--.--.--.--.--.--.--.--.--.--.--.--.--.--.--.--.--.--.--.--.--.--.--.--.--.--.--.--.--.--.--.--.--
\subsubsection{Corrections to the model of Paper I}
%--.--.--.--.--.--.--.--.--.--.--.--.--.--.--.--.--.--.--.--.--.--.--.--.--.--.--.--.--.--.--.--.--.--.--.--

In Paper I, the photodesorption rates were underestimated, as dust extinction was not taken into account. 
Also, a fixed grain size was assumed in the calculation of the cation-grain recombination rates.
However, these only have a minor influence on the gas-phase abundances in the outer wind.
Additionally, CO self-shielding was erroneously not taken into account \citep{VandeSande2018Err}.
Although this significantly affects the size of the CO envelope, the levels of depletion of the parent species are not affected.
The formation of certain minor gas-phase daughter species on the grain surface is affected by the inclusion of CO self-shielding, though this does not have a significant impact on the overall outflow chemistry.
Despite the minor effects of the dust extinction on the photodesorption rates and the inclusion of CO self-shielding on the gas-phase chemistry, we include both effects here for completeness.
Moreover, in Paper I only a single GSD was considered, and we cannot confirm that the effects would be minor if the GSD is changed as is investigated here.

%--.--.--.--.--.--.--.--.--.--.--.--.--.--.--.--.--.--.--.--.--.--.--.--.--.--.--.--.--.--.--.--.--.--.--.--
\subsubsection{Dust-grain size distribution}			\label{subsubsect:gsd}
%--.--.--.--.--.--.--.--.--.--.--.--.--.--.--.--.--.--.--.--.--.--.--.--.--.--.--.--.--.--.--.--.--.--.--.--

{The dust grains in the chemical model are assumed to be compact spherical grains. 
Note that this is different to the grains assumed during continuum radiative transfer modelling, which are used only to extract the dust temperature profiles.
We assume that they} have formed before the start of the model (the region within 20 R$_*$) and follow an MRN-like GSD. 
The dust grain number density is given by
\begin{equation} \label{eq:ndust}
    n_\mathrm{dust} = C \int_{a_\mathrm{min}}^{a_\mathrm{max}} a^{\eta} da \ \ \ \mathrm{cm^{-3}},
\end{equation}
{where $a$ is the radius of the spherical grains}, $a_\mathrm{min}$ and $a_\mathrm{max}$ are the minimum and maximum grain size, and $\eta$ is the exponent of the GSD.
In the canonical MRN distribution, used in Paper I, the parameters are $a_\mathrm{min} = 5 \times 10^{-7}$ cm, $a_\mathrm{max} = 2.5 \times 10^{-5}$ cm and $\eta = -3.5$.

The constant of proportionality, $C$, is derived from the total dust mass density within the outflow, $M$:
\begin{equation}		\label{eq:totaldustmass}
    M = C \int_{a_\mathrm{min}}^{a_\mathrm{max}} a^{\eta} \rho_\mathrm{dust,bulk} \left( \frac{4\pi}{3} a^3 \right) da = \psi \ \rho_\mathrm{gas} \ \ \  \mathrm{g\ cm^{-3}},
\end{equation}
where $\rho_\mathrm{dust,bulk}$ is the bulk density of the dust grains [g cm$^{-3}$], $\rho_\mathrm{gas}$ is the gas mass density [g cm$^{-3}$], and $\psi$ is the dust-to-gas mass ratio.
Together with an assumed value of $\psi$, the constant of proportionality $C$ can be calculated.
We have that 
\begin{equation}
	C = \frac{3(4+\eta)\ \psi\ \rho_\mathrm{gas}}{4 \pi {\rho_\mathrm{dust,bulk}} \left(a_\mathrm{max}^{4+\eta} - a_\mathrm{min}^{4+\eta}\right)}.
\end{equation}
This enables us to calculate the dust grain number density (Eq. \ref{eq:ndust}),
the number density of dust grain surface sites,
\begin{equation}        \label{eq:NS}
    N_s = C\ n_s \int_{a_\mathrm{min}}^{a_\mathrm{max}} a^{\eta} \left(4 \pi a^2 \right) da \ \ \ \mathrm{cm^{-3}},
\end{equation}
where $n_s$ is the density of surface sites on the grain [cm$^{-2}$] (see Table \ref{table:dustparams}), and the average dust grain cross section per unit volume,
\begin{equation}    
    \sigma_\mathrm{dust} = C \int_{a_\mathrm{min}}^{a_\mathrm{max}} a^{\eta} \left( \pi a^2 \right) da\ \ \  \mathrm{cm^{-1}},
\end{equation}
which can be written as
\begin{equation}    \label{eq:sigmadust}
    \sigma_\mathrm{dust} = S\ \frac{4+\eta}{3+\eta}\ \frac{a_\mathrm{max}^{3+\eta} - a_\mathrm{min}^{3+\eta}}{a_\mathrm{max}^{4+\eta} - a_\mathrm{min}^{4+\eta}}\ \ \  \mathrm{cm^{-1}},
\end{equation}
with $S$ a constant specific to the outflow given by
\begin{equation}    \label{eq:S}
    S = \frac{3\ \psi\ \rho_\mathrm{gas}}{4 \pi\ {\rho_\mathrm{dust,bulk}}}.
\end{equation}
The parameter $\sigma_\mathrm{dust}/S$ describes the average dust grain cross section per unit volume in an outflow-independent way.
For the canonical MRN distribution, we have that $\sigma_\mathrm{dust}/S = 2.83 \times 10^5$ cm$^{-1}$.
We will use this parameter to characterise the influence of the specific GSD on the chemistry throughout the outflow, as it comprises all the distribution-specific information on the dust.
%{Note that the number density of grain surface sites (Eq. \ref{eq:NS}) can be written in terms of $\sigma_\mathrm{dust}/S$ as well, so that we have $N_s = 4\ n_s\ \sigma_\mathrm{dust}$.}

\begin{table*}
	\centering
	\caption{Rotational transitions in the ground vibrational state used in our radiative transfer models, together with their frequencies and upper energy levels. 
	}
	\label{table:lines}
%	\resizebox{1.0\columnwidth}{!}{%
	\begin{tabular}{l l l l l l l l l} % four columns, alignment for each
		\hline
	\multicolumn{3}{c}{SiO} & \multicolumn{3}{c}{SiS} & \multicolumn{3}{c}{HCN}\\
		 \cmidrule(lr){1-3}\cmidrule(lr){4-6} \cmidrule(lr){7-9} 
	 Transition & Frequency & $E_\mathrm{upper}$ & Transition & Frequency & $E_\mathrm{upper}$ & Transition & Frequency & $E_\mathrm{upper}$   \\	 
		 \cmidrule(lr){1-3}\cmidrule(lr){4-6} \cmidrule(lr){7-9} 
	 {$J = 2-1$} 	& {43.42 GHz} & {2.084 K} 		& $J = 6-5$ & 108.9 GHz & 18.17 K 	& $J=1-0$ & 88.63 GHz & 4.267 K \\
	 $J = 5-4$ 	& 217.1 GHz & 31.04 K 		& $J=13-12$ & 236.0 GHz & 78.73 K 	& $J=3-2$ & 265.9 GHz & 25.34 K \\
	 $J = 8-7$ 	& 347.3 GHz & 74.50 K 		& $J=19-18$ & 344.8 GHz & 164.4 K 	& $J=8-7$ & 708.9 GHz & 152.0 K  \\
	 $J = 21-20$	& 910.8 GHz & 477.8 K 	 	& $J=37-36$ & 670.5 GHz & 607.7 K 	& $J=10-9$ & 886.0 GHz & 232.3 K   \\
		\hline
	\end{tabular}%
%	}
\end{table*}

%--.--.--.--.--.--.--.--.--.--.--.--.--.--.--.--.--.--.--.--.--.--.--.--.--.--.--.--.--.--.--.--.--.--.--.--
\subsubsection{Parameter selection}
%--.--.--.--.--.--.--.--.--.--.--.--.--.--.--.--.--.--.--.--.--.--.--.--.--.--.--.--.--.--.--.--.--.--.--.--

Table \ref{table:dustparams} lists all GSD parameters used for the models in this paper. 
The parameters $a_\mathrm{min}$, $a_\mathrm{max}$, and $\eta$ are allowed to deviate from their canonical MRN-values (i.e., $a_\mathrm{max} = 2.5 \times 10^{-5}$ cm, $a_\mathrm{min} = 5 \times 10^{-7}$ cm, and $\eta = -3.5$) and are treated as free parameters, while maintaining $a_\mathrm{min} < a_\mathrm{max}$.
\citet{Li2001a} found that roughly 10\% of interstellar Si could be in ultrasmall dust grains with $a \lesssim 15 \times 10^{-8}$ cm.
The smallest possible grains in our models therefore correspond to such ultrasmall dust grains.
Fig. \ref{fig:sigma} shows the variation of $\sigma_\mathrm{dust}/S$ within the selected parameter range.
There are clear degeneracies.
Large values of $\sigma_\mathrm{dust}/S$ can correspond either to small values of $a_\mathrm{min}$ and $a_\mathrm{max}$ and/or to small values of $\eta$, i.e., a steep slope of the GSD.
The value of $a_\mathrm{max}$ cannot be constrained for these GSDs.
For larger values of $\eta$, the value of $\sigma_\mathrm{dust}/S$ is constrained by $a_\mathrm{max}$ rather than $a_\mathrm{min}$.
Pinpointing a specific GSD based on the value of $\sigma_\mathrm{dust}/S$ is impossible.
However, meaningful constraints on whether the grains are predominantly large, or predominantly small, can be made.

%The ratio $a_\mathrm{max}/a_\mathrm{min}$ is a proxy for the efficiency of grain growth in the inner wind, the region before the start of our model at 20 $R_*$; a smaller ratio corresponds to more efficient grain growth in this region.
For values of $\eta < -3.5$, the GSD is skewed to smaller grains than the canonical MRN distribution.
Similarly, for $\eta > -3.5$, the GSD is skewed to larger grains.
As AGB star dust is thought to be large as it enters the ISM \cite[based on observations and theoretical modelling;][]{Groenewegen1997,Winters1997,Gauger1999,DellAgli2017,Nanni2018}, we allow for $\eta > 0$.
This corresponds to more large grains than small grains in the distribution and accounts for the possibility of efficient grain growth throughout the outflow.

%-------------------------------------------------------------------------------------------------------------
\subsection{Radiative transfer modelling}			\label{subsect:meth:rt}
%-------------------------------------------------------------------------------------------------------------

To convert abundance distributions into observable spectra, we use the 1D accelerated lambda iteration method radiative transfer code, ALI.
This code has been used extensively to model a variety of molecules in AGB outflows, e.g. by \citet{Rybicki1991}, \citet{Maercker2008}, \citet{Schoier2011}, and \citet{Danilovich2016}, who describe the code in detail.
We assume a spherically symmetric outflow described by the physical parameters of Table \ref{table:modelparams} and the corresponding dust temperature profiles, as retrieved in Paper I.
The resultant molecular envelope model is then ray-traced assuming generic telescopes with half-power beam-widths of 10\arcsec\ for all synthetic observations at all frequencies. 
The distance to the star is taken to be 500 pc.

The modelled molecular lines are listed in Table \ref{table:lines}.
The different rotational transitions probe different regions of the outflow, where higher energy transitions probe regions closer to the star. 
All transitions lie within the frequency range of current mm and sub-mm telescopes (e.g., APEX, ALMA, SOFIA).
The species SiO, SiS and HCN are depleted in both the O-rich and C-rich outflows, thanks to their high binding energies (3500 K for SiO, 3800 K for SiS, and 3610 K for HCN).
{The collisional rates used for HCN-H$_2$ are scaled from the HCN-He rates of \citet{Dumouchel2010}.
For the SiO-H$_2$ collisional rates, we used scaled SiO-He rates of \citet{dayou2006}, extrapolated by \citet{Schoier2005}.
For SiS-H$_2$ we adopt the SiO-H$_2$ collisional rates.}
Although H$_2$O has a larger binding energy (4880 K) and is therefore very sensitive to dust-gas interactions, we do not use it for radiative transfer modelling.
Cool H$_2$O from the intermediate and outer CSE is only observable using space telescopes, e.g., the retired \textit{Herschel} Space Observatory, due to the presence of water in the atmosphere.

\begin{figure*}
 \includegraphics[width=0.85\textwidth]{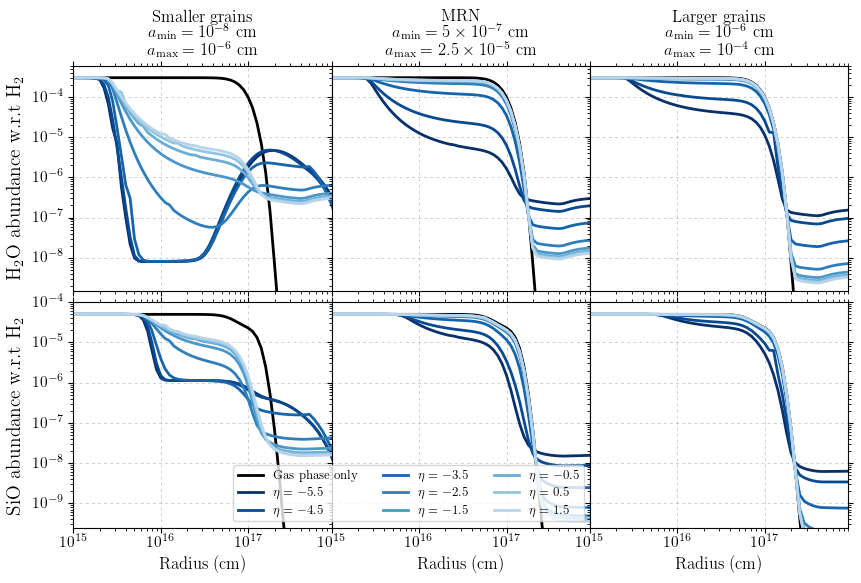}
 \caption{H$_2$O abundance profiles (upper panels) and SiO abundance profiles (lower panels) in an O-rich outflow with melilite dust, characterised by $\dot{M} = 10^{-5} \ \mathrm{M}_\odot\ \mathrm{yr}^{-1}$, $v_\infty$ = 15 km s$^{-1}$, and $v_\mathrm{drift}$ = 10 km s$^{-1}$.
 Middle panel: GSDs with $a_\mathrm{min}$ and $a_\mathrm{max}$ of the MRN distribution, where $\eta = -3.5$ corresponds to the canonical MRN distribution.
 Left and right panels: GSDs with smaller and larger grains than the MRN distribution, respectively.
  }
 \label{fig:depletionH2OSiO}
\end{figure*}

\begin{figure*}
 \includegraphics[width=0.85\textwidth]{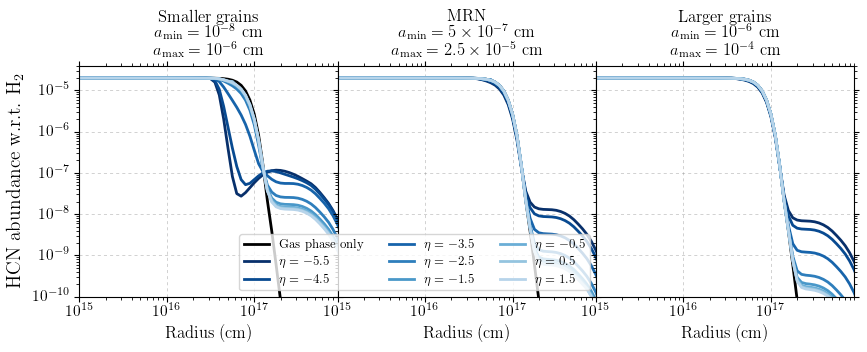}
 \caption{HCN abundance profiles in a C-rich outflow with amorphous carbon dust, characterised by $\dot{M} = 10^{-5} \ \mathrm{M}_\odot\ \mathrm{yr}^{-1}$, $v_\infty$ = 15 km s$^{-1}$, and $v_\mathrm{drift}$ = 10 km s$^{-1}$.
 Middle panel: GSDs with $a_\mathrm{min}$ and $a_\mathrm{max}$ of the MRN distribution, where $\eta = -3.5$ corresponds to the canonical MRN distribution.
 Left and right panels: GSDs with smaller and larger grains than the MRN distribution, respectively.
  }
 \label{fig:depletionHCN}
\end{figure*}

%%%%%%%%%%%%%%%%%%%%%%%%%%%%%%%%%%%%%%%%%%%%%%%%%%%%%%%%%%%%%%%%%%%%%%%%%%%%%%%%%%%%%%%%%%%%%%%%%%%%%%%%%%%%%%
\section{Results}				\label{sect:results}
%%%%%%%%%%%%%%%%%%%%%%%%%%%%%%%%%%%%%%%%%%%%%%%%%%%%%%%%%%%%%%%%%%%%%%%%%%%%%%%%%%%%%%%%%%%%%%%%%%%%%%%%%%%%%%

The choice of GSD affects the chemistry throughout the {higher density outflows considered in this paper.
In lower density outflows, the reduced dust-gas interaction does not lead to a significant depletion of gas-phase species (see Paper I).}
The GSD determines the average grain surface area, encoded in the parameter $\sigma_\mathrm{dust}/S$. 
%A larger value of $\sigma_\mathrm{dust}/S$ {also} leads to a larger number of available surface sites onto which gas-phase species can be depleted {(Eq. \ref{eq:NS})}.
This can be achieved through smaller $a_\mathrm{min}$ and $a_\mathrm{max}$, constraining the distribution to smaller grains, as well as a lower value of the slope, $\eta$, which skews the distribution to smaller grains.

In Sect. \ref{subsect:results:abprof}, we show the influence of the GSD on the depletion of gas-phase species.
The resulting decrease in abundance can influence their observable molecular emission lines, which is then described in Sect. \ref{subsect:results:lines}.

\begin{figure*}
 \includegraphics[width=0.9\textwidth]{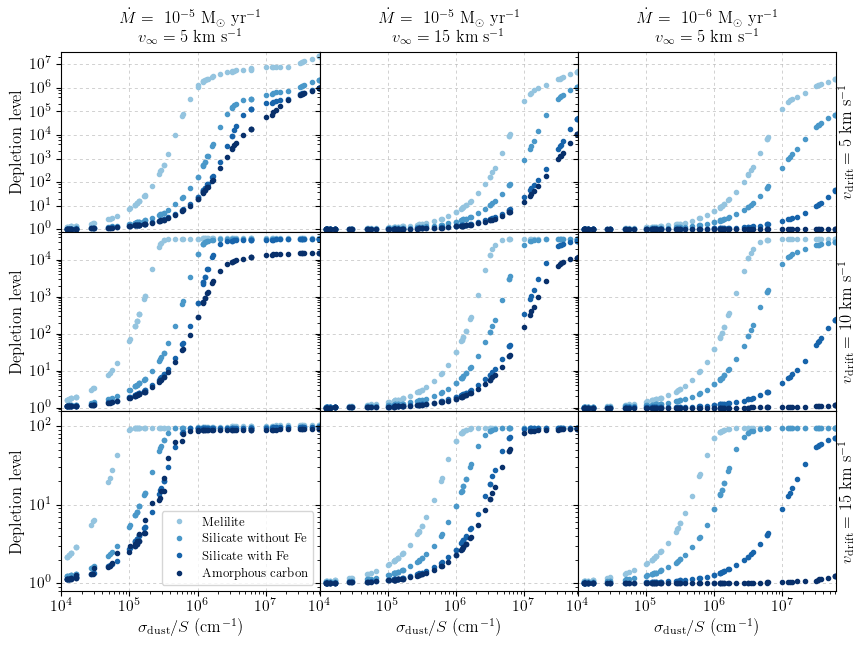}
 \caption{Depletion {level} of the H$_2$O fractional abundance due to dust-gas chemistry, 
{calculated by dividing the abundance obtained from gas-phase chemistry only by the abundance obtained when including dust-gas chemistry, } 
 for outflows with different grain-size distributions, corresponding to different values of $\sigma_\mathrm{dust}/S$ (Eq. \ref{eq:sigmadust}). 
Melilite and silicate with(out) Fe dust correspond to O-rich outflows, amorphous carbon to C-rich outflows.
For $\dot{M} = 10^{-5} \ \mathrm{M}_\odot\ \mathrm{yr}^{-1}$ and $v_\infty$ = 5 km/s, this corresponds to depletion in the region $r < 1 \times 10^{17}$ cm; for $\dot{M} = 10^{-5} \ \mathrm{M}_\odot\ \mathrm{yr}^{-1}$ and $v_\infty$ = 15 km/s to $r < 7 \times 10^{16}$ cm; for $\dot{M} = 10^{-6} \ \mathrm{M}_\odot\ \mathrm{yr}^{-1}$ and $v_\infty$ = 5 km s$^{-1}$ to $r < 2 \times 10^{16}$ cm.
  }
 \label{fig:sigma-H2O}
\end{figure*}

%-------------------------------------------------------------------------------------------------------------
\subsection{Influence on the gas-phase abundance profiles}			\label{subsect:results:abprof}
%-------------------------------------------------------------------------------------------------------------

Fig. \ref{fig:depletionH2OSiO} shows the abundance profiles of H$_2$O and SiO in an O-rich outflow with melilite dust.
Similarly, Fig. \ref{fig:depletionHCN} shows the abundance profiles of HCN in a C-rich outflow with amorphous carbon dust.
These species are selected because of their large initial abundances as parent species (Table \ref{table:Model-Parents}) and their large binding energies, which leads to efficient depletion of the gas-phase species onto the dust.
Both the O-rich and C-rich outflow are characterised by $\dot{M} = 10^{-5} \ \mathrm{M}_\odot\ \mathrm{yr}^{-1}$, $v_\infty$ = 15 km s$^{-1}$, and $v_\mathrm{drift}$ = 10 km s$^{-1}$, i.e., a common high-density outflow \cite[e.g.,][]{Danilovich2015}.
The middle panels show the abundance profiles obtained when assuming the $a_\mathrm{min}$ and $a_\mathrm{max}$ of the MRN distribution (with $\eta = -3.5$ corresponding to the canonical distribution). 
The left and right panels show the results when assuming smaller and larger grains than the MRN distribution in $a_\mathrm{min}$ and $a_\mathrm{max}$, respectively, for different values of $\eta$.
GSDs with smaller grains, be it because of a smaller $a_\mathrm{min}$ and $a_\mathrm{max}$ or a smaller value of $\eta$, lead to a larger depletion of gas-phase species onto the dust.
The effect of dust-gas chemistry on the gas-phase composition of the outflow is determined by other factors as well (see Paper I):
the larger binding energy of H$_2$O leads to a larger depletion compared to SiO, while the warmer amorphous carbon dust gives rise to overall smaller levels of depletion for HCN, despite its similar binding energy to SiO.

The specific GSD also affects the outer wind abundance. 
The increase in gas-phase abundance in this region, {from $\sim 2 \times 10^{17}$ outwards,} is mainly caused by ice mantles sputtering off the dust through collisions with H$_2$, He, CO, and N$_2$, followed by photodesorption.
Since the sputtering rate scales with the average grain surface area, $\sigma_\mathrm{dust}$, GSDs with predominantly smaller grains lead to a larger increase in the outer wind abundance. 
Once returned to the gas phase, the molecules are photodissociated. 
{In outflows with predominantly smaller grains than the MRN distribution (left panel of Fig. \ref{fig:depletionH2OSiO}), the H$_2$O abundance reaches a secondary peak in the outer wind, around $3 \times 10^{17}$ cm, before being photodissociated. 
Because SiO has a larger photodissociation rate than H$_2$O, its abundance does not reach such a peak in the outer wind.}

\begin{figure*}
 \includegraphics[width=0.9\textwidth]{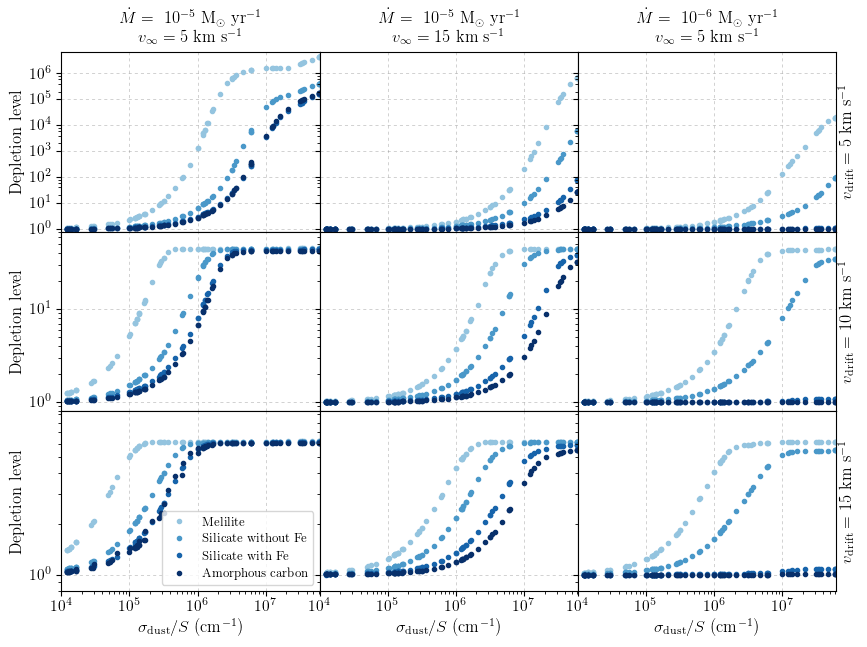}
 \caption{Depletion {level} of the SiO fractional abundance due to dust-gas chemistry, 
{calculated by dividing the abundance obtained from gas-phase chemistry only by the abundance obtained when including dust-gas chemistry, } 
 for outflows with different grain-size distributions, corresponding to different values of $\sigma_\mathrm{dust}/S$ (Eq. \ref{eq:sigmadust}). 
Melilite and silicate with(out) Fe dust correspond to O-rich outflows, amorphous carbon to C-rich outflows.
For $\dot{M} = 10^{-5} \ \mathrm{M}_\odot\ \mathrm{yr}^{-1}$ and $v_\infty$ = 5 km/s, this corresponds to depletion in the region $r < 1 \times 10^{17}$ cm; for $\dot{M} = 10^{-5} \ \mathrm{M}_\odot\ \mathrm{yr}^{-1}$ and $v_\infty$ = 15 km/s to $r < 7 \times 10^{16}$ cm; for $\dot{M} = 10^{-6} \ \mathrm{M}_\odot\ \mathrm{yr}^{-1}$ and $v_\infty$ = 5 km s$^{-1}$ to $r < 2 \times 10^{16}$ cm.
  }
 \label{fig:sigma-SiO}
\end{figure*}

\begin{figure*}
 \includegraphics[width=0.9\textwidth]{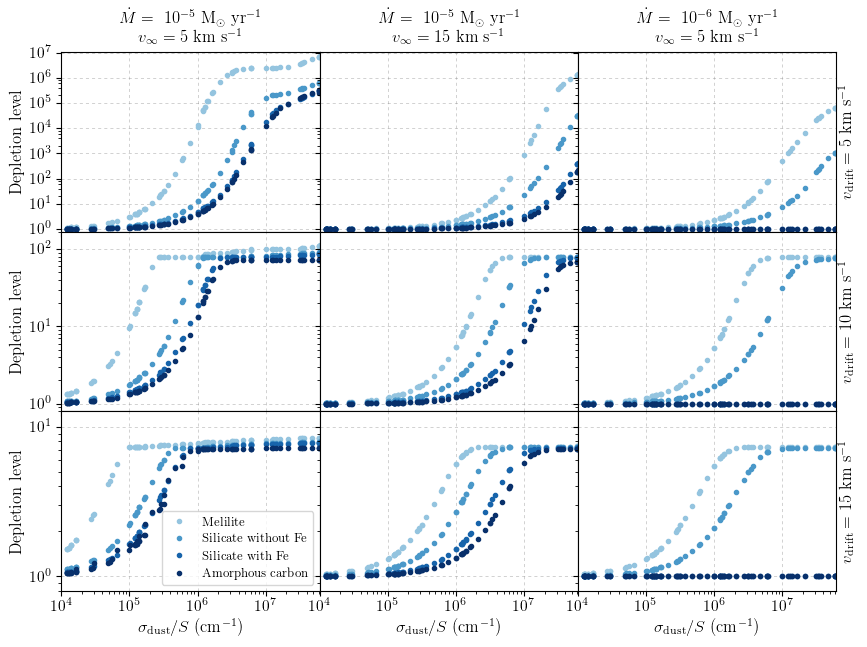}
 \caption{Depletion {level} of the SiS fractional abundance due to dust-gas chemistry, 
{calculated by dividing the abundance obtained from gas-phase chemistry only by the abundance obtained when including dust-gas chemistry, } 
 for outflows with different grain-size distributions, corresponding to different values of $\sigma_\mathrm{dust}/S$ (Eq. \ref{eq:sigmadust}). 
Melilite and silicate with(out) Fe dust correspond to O-rich outflows, amorphous carbon to C-rich outflows.
For $\dot{M} = 10^{-5} \ \mathrm{M}_\odot\ \mathrm{yr}^{-1}$ and $v_\infty$ = 5 km/s, this corresponds to depletion in the region $r < 1 \times 10^{17}$ cm; for $\dot{M} = 10^{-5} \ \mathrm{M}_\odot\ \mathrm{yr}^{-1}$ and $v_\infty$ = 15 km/s to $r < 7 \times 10^{16}$ cm; for $\dot{M} = 10^{-6} \ \mathrm{M}_\odot\ \mathrm{yr}^{-1}$ and $v_\infty$ = 5 km s$^{-1}$ to $r < 2 \times 10^{16}$ cm.
  }
 \label{fig:sigma-SiS}
\end{figure*}

\begin{figure*}
 \includegraphics[width=0.9\textwidth]{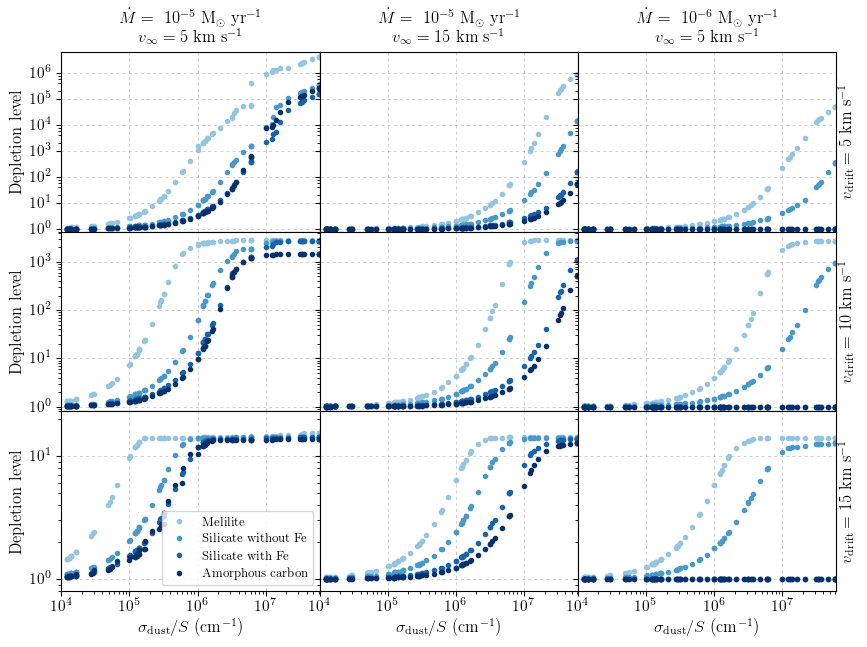}
 \caption{Depletion {level} of the HCN fractional abundance due to dust-gas chemistry, 
{calculated by dividing the abundance obtained from gas-phase chemistry only by the abundance obtained when including dust-gas chemistry, } 
 for outflows with different grain-size distributions, corresponding to different values of $\sigma_\mathrm{dust}/S$ (Eq. \ref{eq:sigmadust}). 
Melilite and silicate with(out) Fe dust correspond to O-rich outflows, amorphous carbon to C-rich outflows.
For $\dot{M} = 10^{-5} \ \mathrm{M}_\odot\ \mathrm{yr}^{-1}$ and $v_\infty$ = 5 km/s, this corresponds to depletion in the region $r < 1 \times 10^{17}$ cm; for $\dot{M} = 10^{-5} \ \mathrm{M}_\odot\ \mathrm{yr}^{-1}$ and $v_\infty$ = 15 km/s to $r < 7 \times 10^{16}$ cm; for $\dot{M} = 10^{-6} \ \mathrm{M}_\odot\ \mathrm{yr}^{-1}$ and $v_\infty$ = 5 km s$^{-1}$ to $r < 2 \times 10^{16}$ cm.
  }
 \label{fig:sigma-HCN}
\end{figure*}

The influence of the GSD on the level of depletion is visualised in Figs. \ref{fig:sigma-H2O}, \ref{fig:sigma-SiO}, \ref{fig:sigma-SiS}, and \ref{fig:sigma-HCN} for H$_2$O, SiO, SiS, and HCN, respectively.
The figures show the depletion {levels} for different outflow densities, drift velocities, and grain types, which correspond to different chemistries.
{The depletion level describes the decrease in abundance by depletion onto dust.
It is calculated by dividing the gas-phase only chemistry abundance by the abundance obtained when including dust-gas chemistry. }
The value of $\sigma_\mathrm{dust}/S$ strongly determines the depletion {level}: a larger average grain surface area leads to a larger level of depletion.
{Depletion} levels off with increasing $\sigma_\mathrm{dust}/S$, for outflows with $v_\mathrm{drift}$ = 10 and 15 km s$^{-1}$ (lower panels).
This marks a balance between accretion, thermal desorption and sputtering.
Outflows with $v_\mathrm{drift}$ = 5 km s$^{-1}$ (upper panels) do not show such a levelling off, as sputtering is not efficient for this drift velocity.
Besides the clear dependence on $\sigma_\mathrm{dust}/S$, other trends, established in Paper I, are visible as well.
The depletion {level increases} with increasing outflow density, as it governs the accretion rate of gas-phase species onto the dust.
Because of a larger sputtering rate, depletion {levels} for $v_\mathrm{drift} = 15$ km s$^{-1}$ are smaller than those for $v_\mathrm{drift} = 10$ km s$^{-1}$.
Finally, colder dust gives rise to a larger depletion of gas-phase species at a certain value of $\sigma_\mathrm{dust}/S$ thanks to the slower thermal desorption rate. 
This leads to C-rich outflows showing {smaller depletion levels} than O-rich outflows, as the amorphous carbon dust is warmer than all O-rich dust species considered (see Paper I).

%-------------------------------------------------------------------------------------------------------------
\subsection{Influence on the molecular line emission}			\label{subsect:results:lines}
%-------------------------------------------------------------------------------------------------------------

\begin{figure*}
 \includegraphics[width=1.0\textwidth]{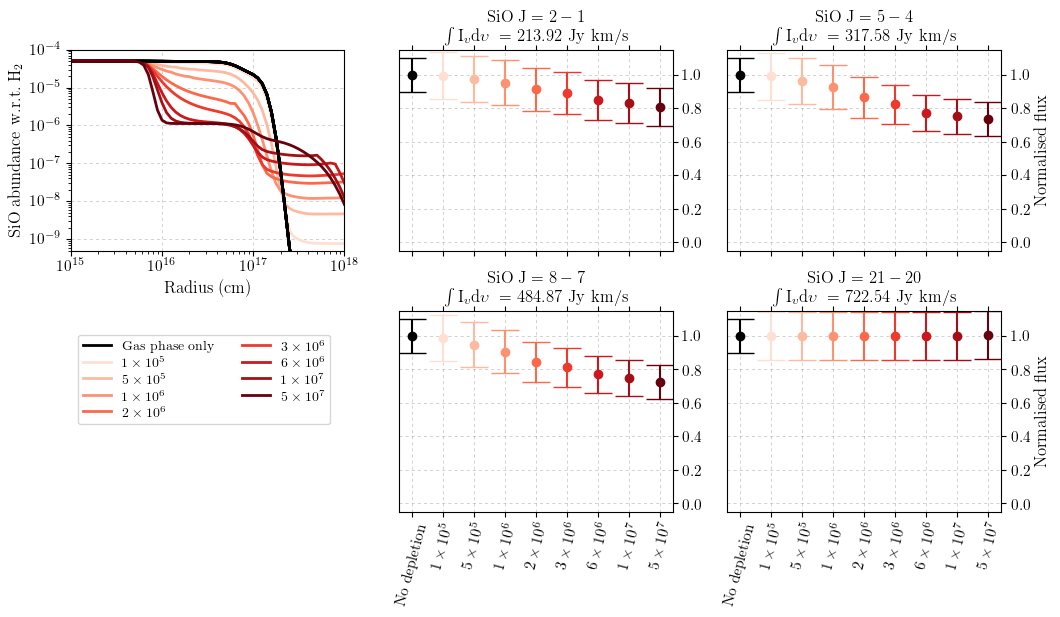}
 \caption{Left panel: SiO abundance in an O-rich outflow with melilite dust, $\dot{M} = 10^{-5} \ \mathrm{M}_\odot\ \mathrm{yr}^{-1}$, $v_\infty$ = 15 km s$^{-1}$, and $v_\mathrm{drift}$ = 10 km s$^{-1}$. 
 Black line: abundance profile without dust-gas chemistry. Coloured lines: abundance profiles with dust-gas chemistry, corresponding to a certain value of $\sigma_\mathrm{dust}/S$ (listed in the legend). 
 Other panels: ratio of the integrated flux of the molecular lines listed in Table \ref{table:lines} when using the corresponding abundance profile over the integrated flux with gas-phase only chemistry. 
 All the line fluxes are normalised to the value of the gas-phase chemistry only model.
 The integrated flux of the gas-phase only molecular lines are given in above the panel.
 The errorbar assumes a 10\% error on the synthetic data.
  }
 \label{fig:RT-SiO}
\end{figure*}

\begin{figure*}
 \includegraphics[width=1.0\textwidth]{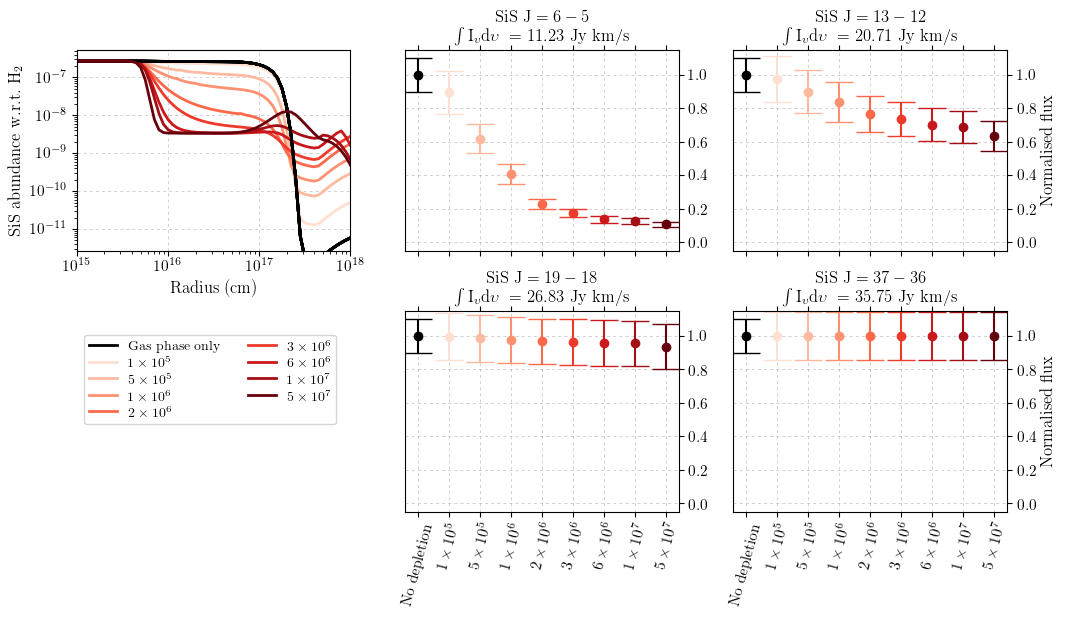}
 \caption{Left: SiS abundance in an O-rich outflow with melilite dust, $\dot{M} = 10^{-5} \ \mathrm{M}_\odot\ \mathrm{yr}^{-1}$, $v_\infty$ = 15 km s$^{-1}$, and $v_\mathrm{drift}$ = 10 km s$^{-1}$. 
 Black line: abundance profile without dust-gas chemistry. Coloured lines: abundance profiles with dust-gas chemistry, corresponding to a certain value of $\sigma_\mathrm{dust}/S$ (listed in the legend). 
 Other panels: ratio of the integrated flux of the molecular lines listed in Table \ref{table:lines} when using the corresponding abundance profile over the integrated flux with gas-phase only chemistry. 
 All the line fluxes are normalised to the value of the gas-phase chemistry only model.
 The integrated flux of the gas-phase only molecular lines are given in above the panel.
 The errorbar assumes a 10\% error on the synthetic data.
  }
 \label{fig:RT-SiS}
\end{figure*}

\begin{figure*}
 \includegraphics[width=1.0\textwidth]{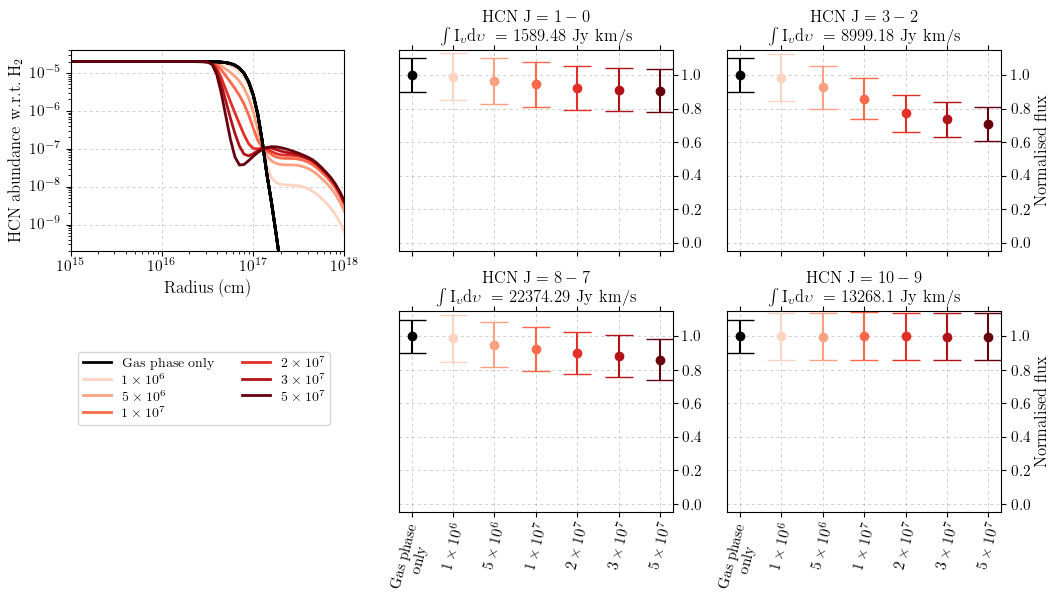}
 \caption{Left: HCN abundance in a C-rich outflow with amorphous carbon dust, $\dot{M} = 10^{-5} \ \mathrm{M}_\odot\ \mathrm{yr}^{-1}$, $v_\infty$ = 15 km s$^{-1}$, and $v_\mathrm{drift}$ = 10 km s$^{-1}$. 
 Black line: abundance profile without dust-gas chemistry. Coloured lines: abundance profiles with dust-gas chemistry, corresponding to a certain value of $\sigma_\mathrm{dust}/S$ (listed in the legend). 
 Other panels: ratio of the integrated flux of the molecular lines listed in Table \ref{table:lines} when using the corresponding abundance profile over the integrated flux with gas-phase only chemistry. 
 All the line fluxes are normalised to the value of the gas-phase chemistry only model.
 The integrated flux of the gas-phase only molecular lines are given in above the panel.
 The errorbar assumes a 10\% error on the synthetic data.
  }
 \label{fig:RT-HCN}
\end{figure*}

Figs. \ref{fig:RT-SiO} and \ref{fig:RT-SiS} show the effect of various GSDs in an O-rich outflow on SiO and SiS line emission, respectively.
Similarly, Fig. \ref{fig:RT-HCN} shows their effect on HCN line emission in a C-rich outflow. 
Both the O-rich and C-rich outflows are characterised by $\dot{M} = 10^{-5}\ \mathrm{M}_\odot\ \mathrm{yr}^{-1}$, $v_\infty$ = 15 km s$^{-1}$, and $v_\mathrm{drift}$ = 10 km s$^{-1}$.
The GSDs are selected according to the level of depletion they cause to the gas-phase abundance, which is reflected in the value of $\sigma_\mathrm{dust}/S$.
The selected depletion ranges from no depletion (only gas-phase chemistry) to the largest level of depletion. 
The abundance profiles used in the radiative transfer modelling are shown in the left panels, together with the corresponding value of $\sigma_\mathrm{dust}/S$.
The right panels show the ratios of the integrated flux of the modelled rotational transitions (Table \ref{table:lines}) to the integrated flux obtained from the gas-phase chemistry only abundance profile, assuming an error of 10\% on the synthetic observations.
The fluxes are normalised to the value from the gas-phase chemistry only model.
For each species, the highest-energy line probes the region before or close to the onset of depletion and is therefore less sensitive to depletion.
The lower-energy lines are affected by depletion, with a lower energy generally corresponding to a larger decrease in integrated flux, as these transitions probe a larger region of the outflow.
The increase in abundance in the outer wind relative to the gas-phase chemistry only models, caused by sputtering and photodesorption, does not significantly contribute to the line fluxes because of the low number density in this region.

For SiO (Fig. \ref{fig:RT-SiO}), the integrated fluxes of the {$J=21-20$ line are indistinguishable}.
The $J=5-4$ and $J=8-7$ lines decrease up to 40\% relative to the gas-phase only chemistry results,
{and the $J=2-1$ line flux decreases up to 20\%.
The decrease in line flux is smaller for the $J=2-1$ line than for the $J=5-4$ line, as the emission region of the lower lines overlaps with the decrease in abundance for all models.
This reduces the overall emission from the $J=2-1$ line, making it less sensitive to depletion effects.}
A clear deviation from the gas-phase only chemistry model can be seen for $\sigma_\mathrm{dust}/S \gtrsim 6 \times 10^6$ cm$^{-1}$.
For SiS (Fig. \ref{fig:RT-SiS}), the $J=37-36$ and $J=19-18$ lines are unaffected by depletion.
The integrated flux of the $J=6-5$ line decreases up to more than 90\% and that of the $J=13-12$ line up to 40\%.
Deviations from the gas-phase only chemistry can be seen for $\sigma_\mathrm{dust}/S \gtrsim 5 \times 10^5$ cm$^{-1}$.
For HCN (Fig. \ref{fig:RT-HCN}), the $J=10-9$, $J=8-7$, and $J=1-0$ lines are unaffected by depletion.
Only the integrated flux of the $J=3-2$ line decreases up to 40\%.
{For $\sigma_\mathrm{dust}/S \gtrsim 2 \times 10^7$ cm$^{-1}$, a deviation from the gas-phase only chemistry model can be retrieved.}

%%%%%%%%%%%%%%%%%%%%%%%%%%%%%%%%%%%%%%%%%%%%%%%%%%%%%%%%%%%%%%%%%%%%%%%%%%%%%%%%%%%%%%%%%%%%%%%%%%%%%%%%%%%%%%
\section{Discussion}				\label{sect:discussion}
%%%%%%%%%%%%%%%%%%%%%%%%%%%%%%%%%%%%%%%%%%%%%%%%%%%%%%%%%%%%%%%%%%%%%%%%%%%%%%%%%%%%%%%%%%%%%%%%%%%%%%%%%%%%%%

The specific GSD included in the chemical kinetics model has a strong influence on the effect of dust-gas chemistry throughout {higher density outflows}.
The MRN-like distribution is described by three parameters: the minimum and maximum grain sizes, $a_\mathrm{min}$ and $a_\mathrm{max}$, and the slope of the distribution $\eta$.
The parameter $\sigma_\mathrm{dust}/S$, the average dust grain cross section, combines all three parameters in an outflow-independent way. 
A larger value of $\sigma_\mathrm{dust}/S$ corresponds to a larger average dust grain surface area.
Therefore, the level of depletion scales with $\sigma_\mathrm{dust}/S$, which is degenerate since a single value of $\sigma_\mathrm{dust}/S$ can correspond to several combinations of $a_\mathrm{min}$, $a_\mathrm{max}$, and $\eta$ (Fig. \ref{fig:sigma}). 
The effect of the GSD on the gas-phase chemistry is in addition to the four main factors established in Paper I, i.e. the density of the outflow, the drift velocity between dust and gas, the temperature of the dust, and the initial composition of the outflow.
Additionally, a different prescription of the GSD, other than a power law, could lead to different results with different degeneracies.

In Sect. \ref{subsect:discussion:retrieval}, we expand on the possibility of retrieving information of the GSD using observations of molecular lines.
In Sect. \ref{subsect:discussion:obs}, we provide a summary of observational results and compare our results to previously obtained observations of the discussed molecular tracers in AGB envelopes.

%-------------------------------------------------------------------------------------------------------------
\subsection{Retrieval of the GSD via line emission}			\label{subsect:discussion:retrieval}
%-------------------------------------------------------------------------------------------------------------

Abundance profiles throughout the outflows of specific AGB stars can be retrieved via radiative transfer modelling when using several molecular transitions that probe different regions of the outflow \cite[e.g.,][]{Decin2010,Agundez2012,VandeSandeRDor,DeBeck2018,Danilovich2019}.
To determine whether depletion is active {in higher density outflows}, causing depleted (lower) abundances in the intermediate wind, it is necessary to include a higher-energy molecular transition that probes the region before the onset of depletion.
{As shown in Figs. \ref{fig:RT-SiO} - \ref{fig:RT-HCN}, depletion influences the relative emission of molecular lines.
By combining high-energy transitions with several lower-energy ones, the abundance profile, as well as any level of depletion, can be retrieved.
The effect on the molecular emission lines is stronger for larger levels of depletion, linked to larger values of $\sigma_\mathrm{dust}/S$, which indicate smaller dust grains. }

As can be seen in Figs. \ref{fig:sigma-H2O} - \ref{fig:sigma-HCN}, the depletion {level} remains constant for $\sigma_\mathrm{dust}/S$ larger than a certain value, which depends on the outflow density, the drift velocity and the type of dust grain.
{This is caused by a balance between accretion, thermal desorption and sputtering.}
{Therefore, if the largest level of depletion is observed for a specific outflow, only the lower limit of the average grain surface area can be retrieved.
For all outflows and dust compositions, the lower limit of $\sigma_\mathrm{dust}/S$ is larger than that of the canonical MRN distribution, which has $\sigma_\mathrm{dust}/S = 2.83 \times 10^5$ cm$^{-1}$.
Retrieving the lower limit is thus also of value, as it indicates that the dust delivered to the ISM is typically smaller than the canonical MRN distribution.
The only exception is  the more extreme outflow characterised by $\dot{M} = 10^{-5} \ \mathrm{M}_\odot\ \mathrm{yr}^{-1}$, $v_\infty$ = 5 km s$^{-1}$, and $v_\mathrm{drift}$ = 5 km s$^{-1}$ containing the coldest (melilite) dust.}
For lower levels of depletion, a more precise value for $\sigma_\mathrm{dust}/S$ can be derived, again assuming that the physical characteristics of the outflow and its dust component are known.
A single value of $\sigma_\mathrm{dust}/S$ can correspond to several combinations of $a_\mathrm{min}$, $a_\mathrm{max}$, and $\eta$, as is shown Fig. \ref{fig:sigma}.
{As a result, the specific GSD cannot be pinpointed due to the degeneracy of $\sigma_\mathrm{dust}/S$ with all three parameters that describe the distribution.}

Nonetheless, information on the specific GSD can still be obtained.
The retrieved (lower limit of) $\sigma_\mathrm{dust}/S$ can be compared with that of the canonical MRN distribution, pointing to more or less dust surface area than the commonly assumed distribution.
More generally, large values of $\sigma_\mathrm{dust}/S$ point towards predominantly small grains, either through small grain sizes (small $a_\mathrm{min}$ and $a_\mathrm{max}$), a steep slope of the distribution (small $\eta$), or a combination of both.
Previous studies \cite[e.g.,][]{Groenewegen1997,Winters1997,Gauger1999,DellAgli2017,Nanni2018} suggest that AGB stars produce large grains ($a \geq 10^{-5}$ cm).
{Assuming $a_\mathrm{max}\gtrsim 10^{-5}$ cm, a larger value for $\sigma_\mathrm{dust}/S$ than that of the canonical MRN distribution is possible by decreasing the value of $\eta$ and/or $a_\mathrm{min}$ (Fig. \ref{fig:sigma}).
Outflows with large dust grains as well as larger values of $\sigma_\mathrm{dust}/S$ therefore also contain more small grains than commonly expected.
This is essential information that complements other approaches to determining the GSD, such as polarisation and SED modelling, which typically assumes the MRN distribution in the optical constants used.}

%-------------------------------------------------------------------------------------------------------------
\subsection{Comparison to observations}			\label{subsect:discussion:obs}
%-------------------------------------------------------------------------------------------------------------

{In order to compare the chemical modelling results to observations, detailed retrieval is necessary.}
Only a few AGB outflows have measured outflow densities, inferred drift velocities, accurate information on the dust composition and temperature, as well as {more precisely} retrieved molecular abundance profiles, {which are crucial. 
Without abundance profiles sampling the entire outflow, depletion levels cannot be obtained.
While several trends of molecular abundance with outflow density have been measured \cite[e.g.,][]{Bujarrabal1989,Sahai1993,Schoier2006a,Massalkhi2019}, only a few depletion levels around specific stars have been observed, as this requires radiative transfer modelling of a set of several both high- and low-energy molecular lines and accompanying SED observations.}

In Sects. \ref{subsubsect:discussion:obs:crich} and \ref{subsubsect:discussion:obs:orich}, we present an overview of observational evidence for depletion of gas-phase molecules in C-rich and O-rich outflows, respectively, and compare them to our models.
For each molecule, we first discuss any observed trends of its abundance with outflow density, followed by any outflow specific depletion levels, and finally a comparison to our models.
Tables \ref{table:CompObs-trend} and \ref{table:CompObs-levels} give a summary of the observed trends with outflow density and the outflow specific depletion levels, respectively.
In Sect. \ref{subsubsect:discussion:retr}, the retrieved GSDs from the observations and their implications are discussed.
All abundances discussed are given with respect to H$_2$.

%--.--.--.--.--.--.--.--.--.--.--.--.--.--.--.--.--.--.--.--.--.--.--.--.--.--.--.--.--.--.--.--.--.--.--.--
\subsubsection{Observations of C-rich outflows}			\label{subsubsect:discussion:obs:crich}
%--.--.--.--.--.--.--.--.--.--.--.--.--.--.--.--.--.--.--.--.--.--.--.--.--.--.--.--.--.--.--.--.--.--.--.--

As IRC+10216 is the most studied AGB star, the majority of observed gas-phase depletion around C-rich stars is found within its outflow, which is characterised by $\dot{M} = 1.5 \times 10^{-5}$ M$_\odot$ yr$^{-1}$ and $v_\infty = 14.5$ km s$^{-1}$ \citep{DeBeck2010}.
As its drift velocity is not know, the retrieved ranges of $\sigma_\mathrm{dust}/S$ depend on the assumed drift velocity within the outflow.

\paragraph*{SiO}
\citet{Schoier2006a} and \citet{Massalkhi2019} retrieved a trend of decreasing SiO molecular abundance with increasing outflow density in a sample of 19 and 25 C-rich AGB stars, respectively, which points towards depletion onto dust grains.
This is successfully predicted by our modelling results.
Additionally, the envelope sizes retrieved by \citet{Schoier2006a} roughly correspond to the SiO envelope sizes in our models.
This indicates that the envelope sizes in low density outflows (smaller than $\sim$ $\dot{M} = 1 \times 10^{-6}$ M$_\odot$ yr$^{-1}$ and $v_\infty = 5$ km s$^{-1}$) are determined by photodissociation, while those in higher density outflows are determined by depletion onto dust (see also Paper I).

For IRC+10216, \citet{Schoier2006b} retrieved that the SiO abundance decreases from $1.5 \times 10^{-6}$ in the inner wind (before $3-8\ R_*$) to $1.7 \times 10^{-7}$.
This corresponds to a depletion level of $\sim$ 9.
\citet{Bujarrabal1989} found depletion levels of 14 and 20 for the V Cyg and S Cep, respectively, based on a relatively simple model.
V Cyg has $\dot{M} = 1 \times 10^{-6}$ M$_\odot$ yr$^{-1}$ and $v_\infty = 10.3$ km s$^{-1}$, while S Cep is characterised by $\dot{M} = 2 \times 10^{-6}$ M$_\odot$ yr$^{-1}$ and $v_\infty = 22.6$ km s$^{-1}$.
As the uncertainties on the observations are large, these depletion levels are uncertain.

Although the SiO depletion takes place before the start of our model, we find that it is depleted in C-rich outflows at around $3 \times 10^{16}$ cm for the outflows with $\dot{M} = 1 \times 10^{-5}$ M$_\odot$ yr$^{-1}$ and $v_\infty = 15$ km s$^{-1}$.
From Fig. \ref{fig:sigma-SiO}, we can derive that the observed depletion level of $\sim$ 9 for SiO around IRC+10216 corresponds to $1 \times 10^7 \lesssim \sigma_\mathrm{dust}/S \lesssim 2 \times 10^7$ cm $^{-1}$.
We do not find significant SiO depletion in C-rich outflows with $\dot{M} = 1 \times 10^{-6}$ M$_\odot$ yr$^{-1}$ and $v_\infty = 5$ km s$^{-1}$, which are similar to the outflows of V Cyg and S Cep, nor lower density outflows.

\paragraph*{SiS}
\citet{Bujarrabal1994} retrieved (upper limits of) SiS abundances within 9 C-rich outflows. They found that these are low compared to equilibrium chemical models, which they suggest could be due to depletion onto dust.
However, \citet{Schoier2007}, \citet{Danilovich2018}, and \citet{Massalkhi2019} found no strong correlation between the SiS abundance and the outflow density in their sample of 14, 4, and 25 C-rich stars, respectively.
We do find that SiS is depleted in higher density outflows.
The scatter in abundance found by \citet{Schoier2007} and \citet{Massalkhi2019} of approximately an order of magnitude for the higher density outflows is consistent with possible levels of depletion (Fig. \ref{fig:sigma-SiS}).
The envelope sizes retrieved by all three studies roughly correspond to the SiS envelope sizes in our models, similar to SiO.
However, \citet{Danilovich2018} noted that smaller envelopes for higher density outflows could also be caused by a larger SiS photodissociation rate. 
 
In IRC+10216, \citet{Agundez2012} found a decrease in SiS abundance around IRC+10216, from $3 \times 10^{-6}$ to $1.3 \times 10^{-6}$.
This corresponds to a depletion level of $\sim$ 2.
The decrease takes place at around $5\ R_*$, i.e. before the start of our model.

Similar to SiO, we find that SiS is depleted in C-rich outflows around $3 \times 10^{16}$ cm for the outflows with $\dot{M} = 1 \times 10^{-5}$ M$_\odot$ yr$^{-1}$ and $v_\infty = 15$ km s$^{-1}$. 
Although the observed depletion is located before the start of our model, we can derive from Fig. \ref{fig:sigma-SiS} that the observed depletion level of $\sim$ 2 corresponds to $3 \times 10^6 \lesssim \sigma_\mathrm{dust}/S \lesssim 10^7$ cm $^{-1}$.
%The value of $\sigma_\mathrm{dust}/S$ depends on the drift velocity within the outflow.

\paragraph*{HCN}

Assuming a scaling law for the HCN envelope size, \citet{Schoier2013} do not find a trend between HCN and outflow density in 25 C-rich outflows.
We do find that HCN is depleted onto dust grains in outflows with  $\dot{M} = 1 \times 10^{-5}$ M$_\odot$ yr$^{-1}$ and $v_\infty = 5$ and $15$ km s$^{-1}$.
\citet{Schoier2013} find a scatter of approximately two order of magnitude in the abundances for higher density outflows, which is consistent with the possible levels of depletion, as shown in Fig. \ref{fig:sigma-HCN}.

Similar to C$_2$H$_2$, \citet{Fonfria2008} found that the HCN abundance around IRC+10216 increases from the photosphere to the inner dust formation zone from $1.23 \times 10^{-5}$ to $4.5 \times 10^{-5}$.
Around 22 $R_*$, they found a decrease in abundance to $2 \times 10^{-5}$, corresponding to a depletion level of 2.25. 

Comparing the retrieved depletion level of $\sim 2.25$ for IRC+10216 to our results, we find from Fig. \ref{fig:sigma-HCN} that it corresponds to $5 \times 10^6 \lesssim \sigma_\mathrm{dust}/S \lesssim 10^7$ cm $^{-1}$.

\paragraph*{CS}
\citet{Danilovich2018} detected CS in a sample of 7 C-rich stars and did not find a clear correlation between abundance and outflow density.
\citet{Massalkhi2019} found that the CS abundance decreases with increasing outflow density in their sample of 25 C-rich stars.
They attribute this to either efficient incorporation into dust in the inner wind or condensation onto the dust further out in the outflow.
The envelope sizes of both studies roughly correspond to those expected by photodissocation.

\citet{Keady1993} found evidence of CS depletion around IRC+10216. 
They found an abundance of $4 \times 10^{-6}$ in the region close to the star ($r \leq 12\ R_*$) and a lower abundance of $1.2 \times 10^{-7}$ further out in the outflow ($r \geq 500\ R_*$), corresponding to a depletion level of $\sim$ 33.
Based on more sensitive data, \citet{Agundez2012} found that the decrease in abundance takes place in the inner wind, around 5 $R_*$. 
They found a decrease in abundance from $4 \times 10^{-6}$ to $7 \times 10^{-7}$, which corresponds to a depletion level of $\sim$ 6. 

Our chemical model is not sensitive to any depletion in the inner wind because it starts at $10^{15}$ cm $\approx 20\ R_*$, i.e., after the inner wind region. 
We are therefore not able to directly compare to the observed depletions of CS in IRC+10216.
In Paper I, we found that CS is only significantly depleted onto dust after $10^{15}$ cm the binding energy of CS is set to 3200 K rather than 1900 K \citep{Garrod2006}, as calculated by \citet{Wakelam2017} for water ice surfaces, forming an ice mantle rather than contributing to dust formation.
With this higher binding energy, the trend observed by \citet{Massalkhi2019} could be due to condensation onto dust rather than incorporation into dust.

\paragraph*{C$_2$H$_2$}
\citet{Fonfria2008} retrieved an increase in C$_2$H$_2$ abundance around IRC+10216 from the photosphere to the inner dust formation zone, around $5\ R_*$, from $7.5 \times 10^{-6}$ to $8.0 \times 10^{-5}$. 
The increase is compatible with equilibrium chemical models as well as non-equilibrium shock chemical models \citep{Cherchneff2006}.

Since our models do not cover the innermost region of the outflow, we cannot comment on the increase in C$_2$H$_2$ abundance before $5\ R_*$, which requires a specific treatment of the dynamic conditions of the inner wind.
We do not find that C$_2$H$_2$ is depleted after $10^{15}$ cm (see Paper I).

In order to compare to the depletion of CS and SiS, and  the increase in C$_2$H$_2$ in the inner wind of IRC+10216, a more comprehensive model that includes dust condensation and is hence valid in the inner wind. 
This is beyond the scope of the current paper.

\begin{table}
	\caption{Overview of observed trends of gas-phase abundance with outflow density of the molecules discussed in Sects. \ref{subsubsect:discussion:obs:crich} and \ref{subsubsect:discussion:obs:orich}.
} 
    \centering
    \begin{tabular}{l c c}
    \hline  
     	 & \multicolumn{2}{c}{Carbon-rich outflows}  \\  
%    \cline{1-6} 
	\hline
%    \noalign{\smallskip}
     &  Trend with  outflow density  & Ref.	\\
    \cline{2-3}
    \noalign{\smallskip}
	\textbf{CS}			& Yes						& (1)		\\    
    \noalign{\smallskip}
	\textbf{SiO}			& Yes						& (1,2)		\\    
    \noalign{\smallskip}
	\textbf{SiS}			& No strong correlation		& (1,3,4)	\\    
%    \noalign{\smallskip}
%	\textbf{C$_2$H$_2$}	&  -							&  			\\       
    \noalign{\smallskip}
	\textbf{HCN}			& No							& (5)		\\                
    \hline 
%%%%%%%%%%%%%%%%%%%%%%%%%%%%%%%%%%%%%%%%%%%%%%%%%%%%
     	 & \multicolumn{2}{c}{Oxygen-rich outflows}  \\  
	\hline
    &   Trend with  outflow density  & Ref.	\\
    \cline{2-3}
    \noalign{\smallskip}
	\textbf{CS}			& No						& (4)		\\    
    \noalign{\smallskip}
	\textbf{SiO}			&  Yes						& (6)		\\    
    \noalign{\smallskip}
	\textbf{SiS}			&  Tentative						& (4,7,8)		\\    
    \noalign{\smallskip}
	\textbf{HCN}			&  No						& (5)		\\    
%    \noalign{\smallskip}
%	\textbf{H$_2$O}			 & -						& 		\\    
	\hline
    \end{tabular}%
    \\
    \footnotesize
    { {{References.}} (1) \citet{Massalkhi2019}; (2) \citet{Schoier2006a}; (3) \citet{Schoier2007};
    (4) \citet{Danilovich2018};  (5) \citet{Schoier2013}; (6) \citet{GonzalezDelgado2003}; (7) \citet{Schoier2007};
    (8) \citet{Danilovich2019}.
    }    
    \label{table:CompObs-trend}    
\end{table}

\begin{table*}
	\caption{Overview of observational evidence for depletion of gas-phase molecules in specific outflows of the molecules discussed in Sects. \ref{subsubsect:discussion:obs:crich} (upper panel) and \ref{subsubsect:discussion:obs:orich} (lower panel).
	The mass-loss rate, $\dot{M}$ (M$_\odot$ yr$^{-1}$), and expansion velocity, $v_\infty$ (km s$^{-1}$) of each object are listed, together with the retrieved depletion level. 
	The estimated $\sigma_\mathrm{dust}/S$ (cm$^{-1}$) is listed for the appropriate depletion levels, excluding depletion taking place before the start of the model and depletion levels with large uncertainty.
} 
    \centering
    \begin{tabular}{l c c c c c c}
    \hline  
     	 & \multicolumn{6}{c}{Carbon-rich outflows}  \\  
%    \cline{1-6} 
	\hline
%    \noalign{\smallskip}
      & Source & $\dot{M}$ & $v_\infty$ &  Depletion level   & Estimated $\sigma_\mathrm{dust}/S$	& Ref.\\
    \cline{2-7}	
    \noalign{\smallskip}
	\textbf{SiO}			& IRC+10216 & $1.5 \times 10^{-5}$ & 14.5  & $\sim$ 9		 & $1 \times 10^7 - 2 \times 10^7$ & (1)\\    
						& V Cyg & $1 \times 10^{-6}$ & 10.3  & 14		 & -   & (2)\\    
						& S Cep & $2 \times 10^{-6}$ & 22.6  & 20		 & - & (2)\\    
    \noalign{\smallskip}
	\textbf{SiS}			& IRC+10216 & $1.5 \times 10^{-5}$ & 14.5 & $\sim$ 2		 & $3 \times 10^6- 10^7$ & (3) \\    
    \noalign{\smallskip}
	\textbf{HCN}			& IRC+10216 & $1.5 \times 10^{-5}$ & 14.5 & 2.25	& $5 \times 10^6- 10^7$  &(4)\\                
    \noalign{\smallskip}
	\textbf{CS}			& IRC+10216	& $1.5 \times 10^{-5}$ & 14.5 & $\sim$ 33  & -  & (5) \\    
						& IRC+10216	& $1.5 \times 10^{-5}$ & 14.5 & $\sim$ 6		& - & (3) \\    
    \noalign{\smallskip}
	\textbf{C$_2$H$_2$}	& IRC+10216 & $1.5 \times 10^{-5}$ & 14.5  & \textit{Increase} by $\sim$ 10 	& -	& (4)	\\       
    \hline 
%%%%%%%%%%%%%%%%%%%%%%%%%%%%%%%%%%%%%%%%%%%%%%%%%%%%
     	 & \multicolumn{6}{c}{Oxygen-rich outflows}  \\  
	\hline
      & Source & $\dot{M}$ & $v_\infty$ &  Depletion level   & Estimated $\sigma_\mathrm{dust}/S$	& Ref.\\
    \cline{2-7}	
%    \noalign{\smallskip}
%	\textbf{CS}			& -	& -	& - & -	& - &  \\    
    \noalign{\smallskip}
	\textbf{SiO}			& R Dor		& $1.2 \times 10^{-7}$ & 5.3 & $\sim$ 13 	& - &(6) \\    
						& L$_2$ Pup	& $2.7 \times 10^{-8}$ & 2.1 & $\sim$ 13 	& - &(6) \\    
						& IK Tau 	& $8 \times 10^{-6}$ & 17.7 & $\sim$ 40		& $\sim\ 5 \times 10^6$ &(7) 	\\    
    \noalign{\smallskip}
	\textbf{SiS}			& IK Tau		& $8 \times 10^{-6}$ & 17.7  & $\sim$ 1375	& $\sim 10^7$ 	& (7)\\    
%    \noalign{\smallskip}
%	\textbf{HCN}			& -	& -	& - & -	& - &  \\    
    \noalign{\smallskip}
	\textbf{H$_2$O}		& OH 127.8+0.0 & $2 \times 10^{-5} - 1 \times 10^{-4}$ & 12.5	& $\sim$ 2	&	 $2 \times 10^5 - 5 \times 10^6$& (8) 	\\    
	\hline
    \end{tabular}%
    \\
    \footnotesize
    { {{References.}}     (1) \citet{Schoier2006b}; (2) \citet{Bujarrabal1989}; (3) \citet{Agundez2012}; (4) \citet{Fonfria2008};
    (5) \citet{Keady1993}; (6) \citet{Schoier2004}; (7) \citet{Decin2010b}; (7) \citet{Lombaert2013}.   
    }    
    \label{table:CompObs-levels}    
\end{table*}

%--.--.--.--.--.--.--.--.--.--.--.--.--.--.--.--.--.--.--.--.--.--.--.--.--.--.--.--.--.--.--.--.--.--.--.--
\subsubsection{{Observations of O-rich outflows}}			\label{subsubsect:discussion:obs:orich}
%--.--.--.--.--.--.--.--.--.--.--.--.--.--.--.--.--.--.--.--.--.--.--.--.--.--.--.--.--.--.--.--.--.--.--.--

\paragraph*{CS}
\citet{Danilovich2018} determined the CS abundance in 3 O-rich stars with a high outflow density ($\dot{M} \sim 10^{-5}$ M$_\odot$ yr$^{-1}$).
They did not find any evidence of depletion or trend between abundance and outflow density.
CS is not included as a parent species in our model (Table \ref{table:Model-Parents}).
The abundances retrieved by \citet{Danilovich2018} are up to two orders of magnitude larger than its peak abundance as a daughter species in our model.
This suggest that for higher density outflows, CS should be included as a parent species.

\paragraph*{SiO}

\citet{GonzalezDelgado2003} found that SiO abundance decreases with increasing outflow density in $\sim$ 40 O-rich outflows, pointing towards depletion onto dust.
Similar to SiO in C-rich outflows, the trend of abundance with outflow density, as well as the retrieved envelope sizes, corresponds to our results.

\citet{Schoier2004} derived from interferometric observations that the SiO abundance profiles in the low mass-loss rate outflows of R Dor and L$_2$ Pup ($\dot{M} = 1.2 \times 10^{-7}$ and $v_\infty = 5.3$ km s$^{-1}$, and $2.7 \times 10^{-8}$ M$_\odot$ yr$^{-1}$ and $v_\infty = 2.1$ km s$^{-1}$, respectively) are composed of a higher abundance inner component with $4 \times 10^{-5}$, which decreases around $3 \times 10^{15}$ cm to $3 \times 10^{-6}$, corresponding to a depletion level of $\sim$ 13. 
We do not find evidence for depletion in such low mass-loss rates outflows (see Paper I).
Additionally, \citet{VandeSandeRDor} do not find evidence of SiO depletion in R Dor and the outflow of L$_2$ Pup is not spherically symmetric, as assumed by \citet{Schoier2004}, but contains a circumstellar dust disk \citep{Kervella2014} which could influence the interpretation of the observations.
Because of these discrepancies, we do not consider the depletion levels retrieved by \citet{Schoier2004}.
For IK Tau, characterised by $\dot{M} = 8 \times 10^{-6}$ M$_\odot$ yr$^{-1}$, $v_\infty = 17.7$ km s$^{-1}$ and $v_\mathrm{drift} = 4$ km s$^{-1}$, \citet{Decin2010} retrieved a decrease in SiO abundance from $1.6 \times 10^{-5}$ to $4.0 \times 10^{-7}$ around 180 $R_*$, a depletion level of 40.
By assuming that only 1\% of the original SiO abundance is left in the gas-phase after depletion onto dust grains after an outflow specific dust formation radius (around $10^{15}$ cm), \citet{Bujarrabal1989} were able to fit the molecular lines of their sample of 9 O-rich outflows, including IK Tau.
As the initial SiO abundance was taken to be $5 \times 10^{-5}$ for all outflows, the assumed depletion corresponds to a depletion level of 500.
For all outflow densities shown in Fig. \ref{fig:sigma-SiO}, such a large depletion level is only found in outflows with $v_\mathrm{drift}$ = 5 km s$^{-1}$.
By assuming that 85\% of gas-phase SiO is depleted onto dust grains, corresponding to a depletion level of $\sim$ 7, \citet{Verbena2019} were able to model interferometric observations of IK Tau and WX Psc. 
The latter is characterised by $\dot{M} = 1 \times 10^{-5}$ M$_\odot$ yr$^{-1}$ and $v_\infty = 20$ km s$^{-1}$.
Their results indicate that SiO depletion is strongly coupled with the gas acceleration, pointing towards condensation of SiO onto dust grains.
Note that the depletion levels of \citet{Bujarrabal1989} and \citet{Verbena2019} are initial assumptions to the modelling, rather than a results of the retrieval.

The dust around IK Tau is thought to be (a combination of) iron-free silicate dust and corundum \citep{Gobrecht2016,Decin2017}.
The SiO depletion level of $\sim$ 40 retrieved by \citet{Decin2010} then points towards $\sigma_\mathrm{dust}/S \approx 10^7$ cm$^{-1}$ (Fig. \ref{fig:sigma-SiO}).
For IK Tau, the depletion level of 500 assumed by \citet{Bujarrabal1989} corresponds to $\sigma_\mathrm{dust}/S \approx 2 \times 10^7$ cm$^{-1}$.
The depletion level of $\sim$ 7 assumed by \citet{Verbena2019} corresponds to $\sigma_\mathrm{dust}/S \approx 5 \times 10^6$ cm$^{-1}$ for both IK Tau and WX Psc.

\paragraph*{SiS}

\citet{Schoier2007} do not find a correlation between SiS and outflow density in their sample of 8 O-rich outflows.
While \citet{Danilovich2018} do not find a correlation in their sample of 3 high outflow density outflows, \citet{Danilovich2019} do find lower abundances in the lower density outflows of their sample of 3 diverse O-rich outflows.
We do find that SiS is depleted in O-rich outflows.
The retrieved envelope sizes correspond to those of our models, where the envelope size is determined by photodissociation in low density outflows and depletion onto dust grains in high density outflows (see also Paper I).

\citet{Decin2010} found a depletion level of $\sim$ 1375 for SiS around IK Tau, where the abundance decreases from $1.1 \times 10^{-5}$ to $8.0 \times 10^{-9}$ around 120 $R_*$.
\citet{Danilovich2019} found a smaller molecular envelope with a inner abundance a factor of 3 lower for IK Tau, which may still be evidence for depletion because of the uncertainty on its photodissociation rate.
To determine the extent to which SiS is depleted in the outflow of IK Tau, a more accurate photodissociation rate for SiS is necessary.

The depletion level of $\sim$ 1375 retrieved for IK Tau points towards $\sigma_\mathrm{dust}/S \approx 10^7$ cm$^{-1}$ (Figs. \ref{fig:sigma-SiO} and \ref{fig:sigma-SiS}), assuming iron-free silicate dust \citep{Gobrecht2016,Decin2017}.

\paragraph*{HCN}

\citet{Schoier2013} do not find a correlation between the HCN abundance and outflow density in 25 O-rich outflows.
However, we do find that their $e$-folding radii of the abundance profiles for larger outflow densities are smaller than predicted by our gas-phase only models. 
This could point towards depletion onto dust grains.

\citet{VandeSandeRDor} do not find evidence for HCN depletion onto dust grains in R Dor, as is expected by our models for low density outflows (see Paper I).

\paragraph*{H$_2$O}

Water ice has been observed in the outflows of OH/IR stars \citep{Omont1990,Justtanont1992}.
These outflows have a large mass-loss rate, typically $\dot{M} \gtrsim 10^{-5}$ M$_\odot$ yr$^{-1}$.
In Paper I, we found that the depletion levels found in our highest density models correspond to the column densities retrieved by \citet{Sylvester1999}, i.e., $10 - 120 \times 10^{16}$ cm$^{-2}$.
\citet{Lombaert2013} suggest a H$_2$O depletion of 50\% in the outflow of OH 127.8+0.0.
They found a mass-loss rate of $2 \times 10^{-5} \gtrsim \dot{M} \gtrsim 1 \times 10^{-4}$ M$_\odot$ yr$^{-1}$ and $v_\infty = 12.5$ km s $^{-1}$ and a dust component composed of predominantly silicate with iron.
The outflow velocity as well as the drift velocity are not known.
\citet{Maercker2016} retrieved H$_2$O abundances and envelope sizes for four O-rich outflows.
These roughly correspond to the envelope sizes in our model, where the envelope sizes in low density outflows are determined by photodissociation, and those in higher density outflows are determined by depletion onto dust.

From the two most dense outflows in Fig. \ref{fig:sigma-H2O} (left and middle panels), we find that a depletion level of 2 corresponds to $2 \times 10^5 \lesssim \sigma_\mathrm{dust}/S \lesssim 5 \times 10^6$ cm $^{-1}$.
The value of $\sigma_\mathrm{dust}/S$ is uncertain, as it depends on the drift velocity as well as the assumed outflow density.

%--.--.--.--.--.--.--.--.--.--.--.--.--.--.--.--.--.--.--.--.--.--.--.--.--.--.--.--.--.--.--.--.--.--.--.--
\subsubsection{Retrieved GSDs from observations}			\label{subsubsect:discussion:retr}
%--.--.--.--.--.--.--.--.--.--.--.--.--.--.--.--.--.--.--.--.--.--.--.--.--.--.--.--.--.--.--.--.--.--.--.--

%{Observed depletion levels are only available for the outflows of IRC+10216, IK Tau and WX Psc, }\textcolor{orange}{together with a rough estimate for OH 127.8+0.0.}
{Besides indirect evidence of depletion through trends of gas-phase abundance with outflow density, reliable observed depletion levels for specific outflows are only available for IRC+10216, IK Tau and WX Psc, together with a rough estimate for OH 127.8+0.0.}
{For the first three outflows, the average grain surface area of their dust component is found to be larger than that of the canonical MRN distribution.}

{Only the lower limit of the (uncertain) range in average grain surface area of the dust around OH 127.8+0.0 corresponds to that of the canonical MRN value.}
{Moreover, modelling of SEDs points towards large grains ($a \geq 10^{-5}$ cm). 
Assuming $a_\mathrm{max}\gtrsim 10^{-5}$ cm, we find from Fig. \ref{fig:sigma} that the value of $\sigma_\mathrm{dust}/S \approx 10^7$ cm$^{-1}$ corresponds to $a_\mathrm{min} \approx 10^{-8}$ cm for $\eta = -3.5$ or $a_\mathrm{min} \approx 10^{-7}$ cm for $\eta = -4.5$ and $-5.5$. }{This value corresponds to the value of $\sigma_\mathrm{dust}/S$ retrieved for IK Tau from SiO and SiS depletion \citep{Decin2010} and the upper limit for the ranges in $\sigma_\mathrm{dust}/S$ retrieved for IRC+10216 from SiO \citep{Schoier2006b}, SiS \citep{Agundez2012}, and HCN depletion \citep{Fonfria2008}.}
{A value of $\sigma_\mathrm{dust}/S \approx 5 \times 10^6$ cm$^{-1}$, } {which can be retrieved for IK Tau and WX Psc from SiO depletion \citep{Verbena2019}, falls within the ranges in $\sigma_\mathrm{dust}/S$ retrieved for IRC+10216, and is slightly larger than the upper limit for OH 127.8+0.0,} {corresponds to $a_\mathrm{min} \approx 10^{-7}$ cm for $\eta = -3.5$ or $a_\mathrm{min} \approx 5 \times 10^{-6}$ cm for $\eta = -4.5$ and $-5.5$.
The values of $a_\mathrm{min}$ roughly corresponds to the size of ultrasmall silicate grains \citep{Li2001a} or to a typical PAH of $\sim$ 50 carbon atoms \citep{Tielens2005}.}
{According to our modelling, the production of large dust grains is accompanied by the injection of more small grains into the ISM than is expected from the canonical MRN distribution.}

{However, these results are obtained from only} {six} {observed depletion levels around} {four} {AGB stars of up to} {four} {possible molecules.
To better constrain the GSD output of AGB outflows, the retrieval of detailed abundance profiles of several molecules within a single outflow is necessary.}
{Moreover, the depletion levels} retrieved from observations are model dependent results because continuum radiative transfer modelling is performed to retrieve the dust temperature profile, based on a certain dust composition and GSD. This feeds into the retrieval of the abundance profiles via line radiative transfer.
{We do not retrieve the canonical MRN distribution commonly used in continuum radiative transfer modelling from our predicted depletion levels in the outflows of IRC+10216, IK Tau and WX Psc.
Only the lower limit of the uncertain range in $\sigma_\mathrm{dust}$ retrieved for OH 127.8+0.0 corresponds to that of the canonical MRN distribution.}
{Together with the retrieved $\sigma_\mathrm{dust}/S$ values corresponding to the GSD being skewed to smaller grains,} the canonical MRN distribution might not be suitable and the retrieved depletion {levels} should be revisited, including a scrutiny of the assumed GSD.

%%%%%%%%%%%%%%%%%%%%%%%%%%%%%%%%%%%%%%%%%%%%%%%%%%%%%%%%%%%%%%%%%%%%%%%%%%%%%%%%%%%%%%%%%%%%%%%%%%%%%%%%%%%%%%
\section{Conclusions}				\label{sect:conclusions}
%%%%%%%%%%%%%%%%%%%%%%%%%%%%%%%%%%%%%%%%%%%%%%%%%%%%%%%%%%%%%%%%%%%%%%%%%%%%%%%%%%%%%%%%%%%%%%%%%%%%%%%%%%%%%%

We included an MRN-like distribution in the chemical kinetics model of Paper I, allowing the GSD to deviate from the commonly assumed canonical MRN distribution.
The GSD is characterised by three parameters: the minimum and maximum grain sizes, $a_\mathrm{min}$ and $a_\mathrm{max}$, and the slope of the distribution, $\eta$.
The parameter $\sigma_\mathrm{dust}/S$ is a function of all three parameters and describes the average dust grain cross section per unit volume in an outflow-independent way.
We used this parameter as a proxy for the specific GSD.

We find that the level of depletion {seen in high density outflows} depends on $\sigma_\mathrm{dust}/S$, where larger values of $\sigma_\mathrm{dust}/S$ give rise to a larger depletion of gas-phase species.
This is because large values of $\sigma_\mathrm{dust}/S$ correspond to a larger average grain surface area within the outflow.
We demonstrated that depletion levels can be retrieved through radiative transfer modelling of several molecular lines that probe different regions within the outflow.
{The depletion level is linked to a value of $\sigma_\mathrm{dust}/S$ within a specific outflow.
Because of the balance between accretion, thermal desorption and sputtering, the depletion level flattens of with increasing $\sigma_\mathrm{dust}/S$ for outflows with $v_\mathrm{drift} \gtrsim 5$ km s$^{-1}$.
Nonetheless, a lower limit of  $\sigma_\mathrm{dust}/S$ can still be determined for the largest depletion levels.}

Although the exact GSD cannot be retrieved, we can determine whether the GSD contains predominantly large or small grains and compare it to the canonical MRN distribution.
Large values of $\sigma_\mathrm{dust}/S$ are due to small minimum grain sizes and/or steep slopes of the GSD.
{For almost all types of outflow considered here, the lower limit that corresponds to the largest level of depletion is larger than the value corresponding to the canonical MRN distribution.}
{While trends of molecular abundance with outflow density have been measured, not many depletion levels have been retrieved for specific AGB stars.}

{From the limited literature sample, we }{generally} {find values of $\sigma_\mathrm{dust}/S$ larger than that of the canonical MRN distribution.
Previous studies suggest that AGB dust is large ($a \geq 10^{-5}$ cm). 
Assuming this is the maximum grain size, larger values of $\sigma_\mathrm{dust}/S$ can be obtained by decreasing the minimum grain size and/or by using a steeper slope of the GSD.
For the steepest slope considered, the minimum grain size corresponds to that of ultrasmall silicate dust or typical PAHs.
Hence, we find that the formation of large dust grains in AGB outflows is accompanied by more small grains than expected by the canonical MRN distribution.}
{Moreover, despite being commonly assumed in continuum radiative transfer modelling, the canonical MRN distribution might not be suitable for AGB outflows, necessitating a revision of the previously retrieved depletion levels.}

{In order to better constrain the GSD throughout AGB outflows, more depletion levels for different molecules within various outflows need to be retrieved from molecular line observations.
This is possible using current mm and sub-mm telescopes.
Our results serve as a proof of concept that depletion levels of gas-phase molecules in higher density outflows, measured from molecular line observations, can constrain the GSD in AGB outflows.
The extra information on the average dust grain cross section is a crucial component to constraining the dust output of AGB stars to the ISM, especially combined with other observations, such as SEDs and polarised light. }

\section*{Acknowledgements}

MVdS and TD acknowledge support from the Research Foundation Flanders (FWO) through grants 12X6419N and 12N9920N, respectively.
CW acknowledges financial support from the University of Leeds and the Science and Technology Facilities Council of the United Kingdom under grant number ST/R000549/1.
{We thank the anonymous referee for helping to improve the paper.}

%%%%%%%%%%%%%%%%%%%%%%%%%%%%%%%%%%%%%%%%%%%%%%%%%%


\bibliographystyle{mnras}
\bibliography{chemistry}

%%%%%%%%%%%%%%%%%%%%%%%%%%%%%%%%%%%%%%%%%%%%%%%%%%

%%%%%%%%%%%%%%%%% APPENDICES %%%%%%%%%%%%%%%%%%%%%

%\appendix


%%%%%%%%%%%%%%%%%%%%%%%%%%%%%%%%%%%%%%%%%%%%%%%%%%

% Don't change these lines
\bsp	% typesetting comment
\label{lastpage}
\end{document}